%% file: Reference_Governor_Strategies_for_Vehicle_Rollover_Avoidance.tex
\documentclass[journal]{IEEEtran}

\usepackage{url}
\usepackage[printonlyused,nohyperlinks]{acronym} % For the List of Abbreviations. Read acronym.pdf on ctan.tug.org
\usepackage{latexsym}
\usepackage[]{algorithm2e}

\usepackage{ifpdf}
 \ifpdf
 \else
 \fi
\usepackage{cite}
\ifCLASSINFOpdf
  \usepackage[pdftex]{graphicx}
  \graphicspath{{./pdf/}{./jpeg/}{./png/}}
  \DeclareGraphicsExtensions{.pdf,.jpeg,.png}
\else
  \usepackage[dvips]{graphicx}
  \graphicspath{{./eps/}}
  \DeclareGraphicsExtensions{.eps}
\fi
\usepackage[cmex10]{amsmath}
\usepackage{amsfonts}
\usepackage{amssymb}
\usepackage{amsthm}
\usepackage{algorithmic}
\usepackage{array}
\ifCLASSOPTIONcompsoc
  \usepackage[caption=false,font=normalsize,labelfont=sf,textfont=sf]{subfig}
\else
  \usepackage[caption=false,font=footnotesize]{subfig}
\fi
\usepackage{stfloats}
\fnbelowfloat
\theoremstyle{plain}
\newtheorem{theorem}{Theorem}%[chapter]

\theoremstyle{definition}

\theoremstyle{remark}
\newtheorem{remark}[theorem]{Remark}
\newcommand{\argmin}{\arg\!\min}
\newcommand{\argmax}{\arg\!\max}

\begin{document}
\title{Reference Governor Strategies for Vehicle Rollover Avoidance}

\author{Ricardo~Bencatel,
        Anouck~Girard,~\IEEEmembership{Senior Member,~IEEE,}
        and~Ilya~Kolmanovsky,~\IEEEmembership{Fellow,~IEEE}% <-this % stops a space
\thanks{R. Bencatel {\scriptsize (corresponding author)}, A. Girard, and I. Kolmanovsky are with the Department
of Aerospace Engineering, University of Michigan, Ann Arbor,
MI, 48109 USA; e-mail: {ricardo.bencatel@fe.up.pt, \{anouck,ilya\}@umich.edu}.}% <-this % stops a space
\thanks{Manuscript received ??? \#, 2016; revised ??? \#, 2016.}}


\markboth{IEEE Transactions on Control Systems Technology,~Vol.~\#, No.~\#, ???~2016}%
{Reference Governor Strategies for Vehicle Rollover Avoidance}



\maketitle

\begin{abstract}

The paper addresses the problem of vehicle rollover avoidance using reference governors applied to modify the driver steering input in vehicles with an active steering system.
Several reference governor designs are presented and tested with a detailed nonlinear simulation model.
The vehicle dynamics are highly nonlinear for large steering angles, including the conditions where the vehicle approaches a rollover onset, which necessitates reference governor design changes.
Simulation results show that reference governor designs are effective in avoiding rollover.
The results also demonstrate that the controllers are not overly conservative, adjusting the driver steering input only for very high steering angles.

\end{abstract}


% \fi
\IEEEpeerreviewmaketitle

\input{Abbrev}

\input{Introduction}

% Reminder: the "draftcls" or "draftclsnofoot", not "draft", class
% option should be used if it is desired that the figures are to be
% displayed while in draft mode.
%\begin{figure*}[!t]
%\centering
%\subfloat[Case I]{\includegraphics[width=2.5in]{box}%
%\hfil
%\subfloat[Case II]{\includegraphics[width=2.5in]{box}%
%\label{fig_second_case}}
%\caption{Simulation results for the network.}
%\label{fig_sim}
%\end{figure*}

%\begin{table}[!t]
%% increase table row spacing, adjust to taste
%\renewcommand{\arraystretch}{1.3}
% if using array.sty, it might be a good idea to tweak the value of
% \extrarowheight as needed to properly center the text within the cells
%\caption{An Example of a Table}
%\label{table_example}
%\centering
%% Some packages, such as MDW tools, offer better commands for making tables
%% than the plain LaTeX2e tabular which is used here.
%\begin{tabular}{|c||c|}
%\hline
%One & Two\\
%\hline
%Three & Four\\
%\hline
%\end{tabular}
%\end{table}

% Note that, LaTeX2e, unlike IEEE journals/conferences, places
% footnotes above bottom floats. This can be corrected via the
% \fnbelowfloat command of the stfloats package.

%=====================================================================================%
%=====================================================================================%
\section{Vehicle and Constraint Modeling}
\label{sec:Models}

\input{Model_NonlinearCar}

\input{Model_Constraints}

\input{Model_LinearCar}
\input{Model_PerformanceMetrics}

%=====================================================================================%
%=====================================================================================%
\input{Ctrl_RG}
\label{sec:RefGovernors}

%=====================================================================================%
%=====================================================================================%
\section{Simulation Results}
\label{sec:Results}

\input{Results_SimSetup}

\input{Results_RG_Steering}

\input{Results_CtrlComp_NoError}

\input{Results_CtrlComp_EstimError}

%%=====================================================================================%
%\subsection{Differential Braking Command Governor}
%
%Comparison between the reference trajectory and the different command governors:
%\begin{enumerate}
% \item Constraint respect;
% 
% \item Differential braking activation minimization;
% 
% \item Horizontal trajectory;
% 
% \item Relevant lateral dynamics: side slip and roll.
%\end{enumerate}
%
%
%%=====================================================================================%
%\subsection{Active Suspension Command Governor}
%
%Comparison between the reference trajectory and the different command governors:
%\begin{enumerate}
% \item Constraint respect;
% 
% \item Active suspension adjustments minimization;
% 
% \item Horizontal trajectory;
% 
% \item Relevant lateral dynamics: side slip and roll.
%\end{enumerate}
%
%
%%=====================================================================================%
%\subsection{Combined Command Governor}
%
%Comparison between the reference trajectory and the best command governors (probably nonlinear extended) with Euclidean- and 1-norm costs:
%\begin{enumerate}
% \item Constraint respect;
% 
% \item Minimization of the commands adjustments;
% 
% \item Horizontal trajectory;
% 
% \item Relevant lateral dynamics: side slip and roll.
%\end{enumerate}

%=====================================================================================%
%=====================================================================================%
\input{Conclusions}

% or


%\appendices
%\section{}
%\section{}

%=====================================================================================%
%=====================================================================================%
\section*{Acknowledgment}

The authors would like to thank ZF-TRW that partially funded this work, and in particular Dan Milot, Mark Elwell and Chuck Bartlett for their technical support and guidance in the controllers' development process.

\ifCLASSOPTIONcaptionsoff
  \newpage
\fi



\bibliographystyle{IEEEtran}
\bibliography{IEEEabrv,References}
%\bibliography{IEEEabrv,H:/References/References}


%\begin{IEEEbiography}{Ricardo Bencatel}
%\end{IEEEbiography}
%
%\begin{IEEEbiographynophoto}{Anouck Girard}
%% insert where needed to balance the two columns on the last page with
%%\newpage
%
%\begin{IEEEbiographynophoto}{Ilya Kolmanovsky}


%\vfill


\end{document}

%% file: Abbrev.tex
%================================================================================================================================%
%================================================================================================================================%
\section*{Acronyms}

{\small
\begin{acronym}[NHTSA]
 \input{abbr}
\end{acronym}
}

%% file: abbr.tex
\acro{AOA}{Angle-of-Attack}

\acro{AR}{Aspect Ratio}

\acro{AUV}{Autonomous Underwater Vehicle}

\acro{ASV}{Autonomous Surface Vehicle}

\acro{APF}{Adaptive Particle Filter}

\acro{AWA}{Apparent Wind Angle}

\acro{AWS}{Apparent Wind Speed}

\acro{BWA}{B?? Wind Angle}

\acro{CAS}{Collision Avoidance System}

\acro{CG}{Command Governor}

\acro{CM}{Center of Mass}

\acro{cdf}{cumulative distribution function}

\acro{CPCAA}{Close Proximity Collision Avoidance Algorithm}

\acro{CWA}{C?? Wind Angle}

\acro{DPG}{Deconflicting Path Generator}

\acro{DOF}{Degrees-of-Freedom}

\acro{ECG}{Extended Command Governor}

\acro{EKF}{Extended Kalman Filter}

\acro{ERG}{Extended Reference Governor}

\acro{ESC}{Electronic Stability Control}

\acro{ESP}{Electronic Stability Program}

\acro{FCS}{Flight Control System}

\acro{FOL}{First-Order Logic}

\acro{FRC}{Foil Rotation Center}

\acro{GNSS}{Global Navigation Satellite System}

\acro{GP}{Gaussian process}

\acro{GPS}{Global Positioning System}

\acro{GUI}{Graphical User Interface}

\acro{HCS}{Hybrid Control System}

\acro{HIL}{Hardware-in-the-Loop}

\acro{IID}{Independent and Identically Distributed}

\acro{IC}{Internal Combustion}

\acro{IMU}{Inertial Measurement Unit}

\acro{KF}{Kalman Filter}

\acro{LQR}{Linear-Quadratic Regulator}

\acro{LRG}{Linear Reference Governor}

\acro{LTR}{Load Transfer Ratio}

\acro{LWS}{Layer Wind Shear}

\acro{MWA}{Mark Wind Angle}

\acro{MPC}{Model Predictive Controller}

\acro{MPL}{Multi-Point Linearization}

\acro{NM}{nautical miles}

\acro{NHTSA}{National Highway Traffic Safety Administration}

\acro{NRG}{Nonlinear Reference Governor}

\acro{PDA}{Path Deconflicting Algorithm}

\acro{P}{Proportional}

\acro{PD}{Proportional Derivative}

\acro{PI}{Proportional Integrative}

\acro{PID}{Proportional Integrative Derivative}

\acro{pdf}{probability density function}

\acro{PL}{Propositional Logic}

\acro{PF}{Particle Filter}

\acro{QP}{Quadratic Programming}

\acro{RAPF}{Regularized Adaptive Particle Filter}

\acro{RG}{Reference Governor}

\acro{RHC}{Receding Horizon Control}

\acro{RPF}{Regularized Particle Filter}

\acro{RPM}{Rotations Per Minute}

\acro{SCG}{Scalar Command Governor}

\acro{SIL}{Software-in-the-Loop}

\acro{SMC}{Sliding Mode Control}

\acro{SWS}{Surface Wind Shear}

\acro{TCAS}{Traffic Alert and Collision Avoidance System}

\acro{TL}{Temporal Logic}

\acro{TWA}{True Wind Angle}

\acro{TWS}{True Wind Speed}

\acro{UAV}{Unmanned Aerial Vehicle}

\acro{UAS}{Unmanned Aircraft System}

\acro{UI}{User Interface}

\acro{VMG}{Velocity Made Good} 

\acro{VMM}{Velocity to the Mark} 

\acro{VPP}{Velocity Prediction Program} 

\acro{VSC}{Vehicle Stability Control} 

\acro{VTOL}{Vertical Takeoff and Landing} 

\acro{WIP}{Work in Progress}

\acro{ZOH}{Zero Order Hold}

%% file: Introduction.tex
\section{Introduction}
\label{sec:Introduction}

\IEEEPARstart{R}{ollover} is an event where a vehicle's roll angle increases abnormally. %, leading to a state where the angle between the vehicle body vertical axis and the ground vertical axis is larger than 90 degrees. %\emph{Reference?}.
In most cases, this results from a loss of control of the vehicle by its driver.

This work focuses on the design of a supervisory controller that intervenes to avoid such extreme cases of loss of control.
Conversely, in operating conditions considered normal the controller should not intervene, letting the driver commands pass through unaltered.

%\IEEEpubidadjcol

%================================================================================================================================%
\subsection{Problem Statement}
\label{sec:IntroProbStatement}

This paper treats the following problem for a vehicle equipped with an active front steering system.
\emph{Given vehicle dynamics, a control model, and a set of predefined rollover avoidance constraints, find a control law for the steering angle such that the defined constraints are always enforced and the applied steering is as close as possible to that requested by the vehicle's driver.}
%This study treats the problems defined below:
%\begin{itemize}
% \item[P1] Given car dynamics and control model and a set of predefined rollover avoidance constraints, find a control law for the steering angle such that the defined constraints are always respected and the applied steering is as close as possible to that requested by the vehicle driver.
%
% \item[P2] Given a car dynamics and control model and a set of predefined rollover avoidance constraints, find a control law for the differential braking such that the defined constraints are always respected and the applied differential braking is minimized.
% 
% \item[P3] Given a car dynamics and control model and a set of predefined rollover avoidance constraints, find a control law for the active suspension such that the defined constraints are always respected and the active suspension adjustments are minimized.
%
% \item[P4] Given a car dynamics and control model and a set of predefined rollover avoidance constraints, find a control law for the steering angle, differential braking, and active suspension such that the defined constraints are always respected and the applied control variables are as close as possible to those requested by the vehicle driver.
%\end{itemize}

%================================================================================================================================%
\subsection{Motivation}

Between 1991 and 2001 there was an increase in the number of vehicle rollover fatalities when compared to fatalities from general motor vehicle crashes \cite{NHTSA2003Rollover}. 
This has triggered the development of safety test standards and vehicle dynamics control algorithms to decrease vehicle rollover propensity.
Rollover remains one of the major vehicle design and control considerations, in particular, for larger vehicles such as SUVs.

%================================================================================================================================%
\subsection{Background and Notation}

A variety of technologies, including differential braking, active steering, active differentials, active suspension and many others, are already in use or have been proposed to assist the driver in maintaining control of a vehicle.
Since the introduction of the \ac{ESP} \cite{vanZanten2000Car_ESP}, much research has been undertaken on the use of active steering, see e.g., \cite{Ackermann1999Car_ActiveSteering}, to further enhance vehicle driving dynamics.
According to \cite{Carlson2003Car_RolloverPrevention} and \cite{Solmaz2007RolloverPrevention_ActiveSteering}, there is a need to develop driver assistance technologies that are transparent to the driver during normal driving conditions, but would act when needed to recover handling of the vehicle during extreme maneuvers.
The active steering system has been introduced in production by ZF and BMW in 2003 \cite{Koehn2004ActiveSteering_BMW}.

For the rollover protection problem, Solmaz \emph{et al.} \cite{Solmaz2006RolloverPrevention_DifferentialBraking,Solmaz2007RolloverPrevention_ActiveSteering_Conf,Solmaz2007RolloverPrevention_ActiveSteering} develop robust controllers which reduce the \acl{LTR}'s magnitude excursions.
The presented controllers are effective at keeping the \ac{LTR} within the desired constraints.
Their potential drawbacks are that the controller is always active, interfering with the nominal steering input, or is governed by an \emph{ad hoc} activation method.
Furthermore, the controllers were tested with a linear model, whose dynamics may differ significantly from more realistic nonlinear models, in particular for larger steering angles, at which rollover is probable. % might occur

Constrained control methods have evolved significantly in recent years to a stage where they can be applied to vehicle control to enforce pointwise-in-time constraints on vehicle variables, thereby assisting the driver in maintaining control of the vehicle.
Reference \cite{Falcone2008Car_ActiveSteering} is an indication of this trend.
%In the present paper we consider an application of the techniques based on constraint admissible sets \cite{Kolmanovsky1998DisturbanceInvariantSets}, reference governors \cite{Gilbert1999ReferenceGovernors} and Model Predictive Control for handling constraints in vehicles with active steering.

%A special formulation of Model Predictive Control, related to the work \cite{Bemporad1997PredictiveReferenceManagement}, \cite{Gilbert2011CommandGovernor}, will be employed.
%The cost function of Model Predictive Controller and terminal set constraints will be defined in such a way that the steering becomes modified only when necessary to avoid constraint violation.
%The reference governor will be introduced as the simplest kind of such a Model Predictive Controller, corresponding to the decision horizon of one step, and its properties will be compared to a more general Model Predictive Controller with the multi-step decision horizon.
%For the implementation of the Model Predictive Controller, an explicit solution to the associated parametric \ac{QP} problem can be developed \cite{Bemporad2002LQR_ConstrainedSystems}.
%We will show that the complexity of this explicit solution can be significantly reduced by using simply characterized constraint feasible sets.

We employ mostly standard notations throughout.
We use $ \left[ a, b \right] $ to denote an interval (subset of real numbers between $a$ and $b$) for either $ a < b $ or $ a > b $.

%================================================================================================================================%
\subsection{Original Contributions}

%%========= Figure ===============%
%\begin{figure}[!b]
% \centering
% \subfloat[Top view.]{
%  \includegraphics[height = 1.8in]{TopView}
%  \label{fig:TopView}
% }
% \\
% %\hspace{1cm}
%% \centering
%% \subfloat[Side view.]{
%%  \includegraphics[height = 2.25in]{SideView}
%%  \label{fig:SideView}
%% }
% \subfloat[Rear view.]{
%  \includegraphics[height = 1.8in]{RearView}
%  \label{fig:RearView}
% }
% \caption{Vehicle forces diagram.}
% \label{fig:VehicleViews}
%\end{figure}
%%========= Figure ===============%

This paper illustrates the development and the application of reference and extended command governors to \ac{LTR} constraint enforcement and vehicle rollover avoidance.
The reference governor (see \cite{Kolmanovsky2014CommandGovernor} and references therein) is a predictive control scheme that supervises and modifies commands to well-designed closed-loop systems to protect against constraint violation.
In the paper, we consider the application of both linear and nonlinear reference governor design techniques to vehicle rollover protection.
While in the sequel we refer to the reference governors modifying the driver steering input, we note that they can be applied to modify the nominal steering angle generated by another nominal steering controller.

Our linear reference governor design exploits a family of linearizations of the nonlinear vehicle model for different steering angles.
The linearized models are used to predict the vehicle response and to determine admissible steering angles that do not lead to constraint violation.
In an earlier conference paper \cite{Kolmanovsky2009CommandGovernor_Steering}, reference and extended command governor designs for steering angle modification were proposed and validated on a linear vehicle model.
This paper is distinguished by extending the design methodology to include compensation for nonlinear behavior, and validating the design on a comprehensive nonlinear vehicle model that includes nonlinear tire, steering, braking and suspension effects.
The strong nonlinearity of the vehicle dynamics for high steering angles, the conditions where the vehicle is at risk of rolling over, caused the simpler controller presented in \cite{Kolmanovsky2009CommandGovernor_Steering} to produce steering commands that were too conservative.
The \ac{LRG} and \ac{ECG} presented in Section \ref{sec:RefGovernors} compensate for the strong nonlinearities with low interference over the driver input, while maintaining a low computational burden and a high effectiveness in avoiding car wheel lift. % (precursor to rollover).

For comparison, a \ac{NRG} is also developed, which uses a nonlinear vehicle model for prediction of constraint violation and for the onboard optimization.
This online reference governor approach is more computationally demanding but is able to take into account nonlinear model characteristics in the prediction.

The original contributions of this work are summarized as follows:
\begin{enumerate}
% \item This study presents a vehicle model comprehensive enough to include steering, braking and suspension effects, while still simple enough to allow an efficient controller synthesis.
% \item This study presents a comprehensive vehicle model, with steering, braking, and suspension controls, while still simple enough to allow an efficient controller synthesis.
 
 \item This paper demonstrates how Reference and \aclp{ECG} \cite{Kolmanovsky2014CommandGovernor}, both linear and nonlinear, can improve the output adherence to the driver reference, while maintaining the vehicle within the desired constraints.
 
 \item Conservatism, effectiveness, and turning response metrics are defined to evaluate, respectively, the controller command adherence to the driver reference, the constraints enforcement success, and the adherence of the vehicle trajectory in the horizontal plane to that desired by the driver.
 
% \item The presented work shows how steering adjustments, differential braking, and active suspension adjustments can be combined to minimize the overall adjustments.
 
 \item Several methodological extensions to reference governor techniques are developed that can be useful in other applications.
\end{enumerate}

%Contribution 1 lays down the models used to synthesize and test the control laws. % used to solve problems P1, P2, P3, and P4.
%Contribution 2 builds on contribution 1 to set the control strategies, addressing problems P1, P2, P3, and P4.
%Contribution 3 addresses problem P4.

%Contribution 1 defines the control strategies, addressing the problem stated in \ref{sec:IntroProbStatement} and
%Contribution 2 allows the performance evaluation required to demonstrate contribution 1.

%================================================================================================================================%
\subsection{Paper Structure}

%The remainder of the paper is divided in three main sections: Models (sec.~\ref{sec:Models}), Controllers (sec.~\ref{sec:RefGovernors}), and Results (sec.~\ref{sec:Results}).
The paper is organized as follows.
Section~\ref{sec:Models} describes the vehicle models as well as the control constraints and the performance metrics used to evaluate the vehicle dynamic response under different designs.
Section~\ref{sec:RefGovernors} describes reference governors considered in this paper, and includes preliminary performance evaluation to support the controller design decisions.
Section~\ref{sec:Results} illustrates the simulation results obtained with the different \acp{RG} and comments on their comparative performance.
Section~\ref{sec:Conclusions} describes the conclusions of the current work, the current complementary research, and the follow-up work envisioned.

%% file: Model_NonlinearCar.tex
\subsection{Nonlinear Car Model}
\label{sec:Model_NonlinearCar}

The nonlinear vehicle dynamics model is developed following \cite{Zhou2009Car_ActiveSafety_Impacts,Solmaz2007RolloverPrevention_ActiveSteering,Ulsoy2012CarDynamicsControl,Abe2015CarDynamicsControl}.
The model includes a nonlinear model for the tires' friction and the suspension.
The suspension model includes parameters to allow the simulation of a differential active suspension system.

%=====================================================================================%
\subsubsection{Vehicle Body Equations of Motion}
\label{sec:Model_NonlinearCar_BodyEOM}

\begin{figure}[t]
 \centering
 \subfloat[Rear view.]{
  \includegraphics[height = 1.7in]{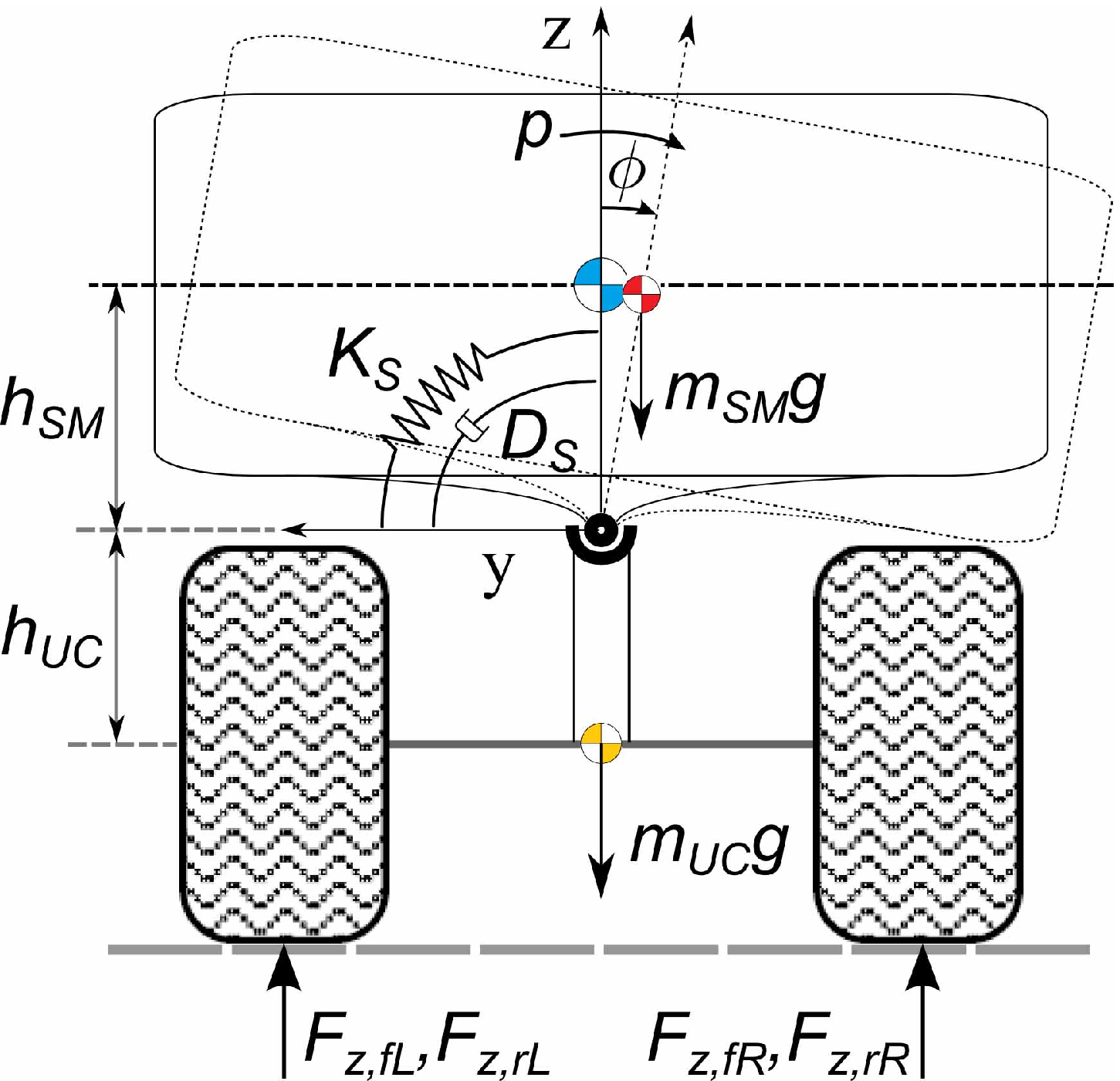}
  \label{fig:RearView}
 }
 \\
 \subfloat[Top view.]{
  \includegraphics[height = 1.5in]{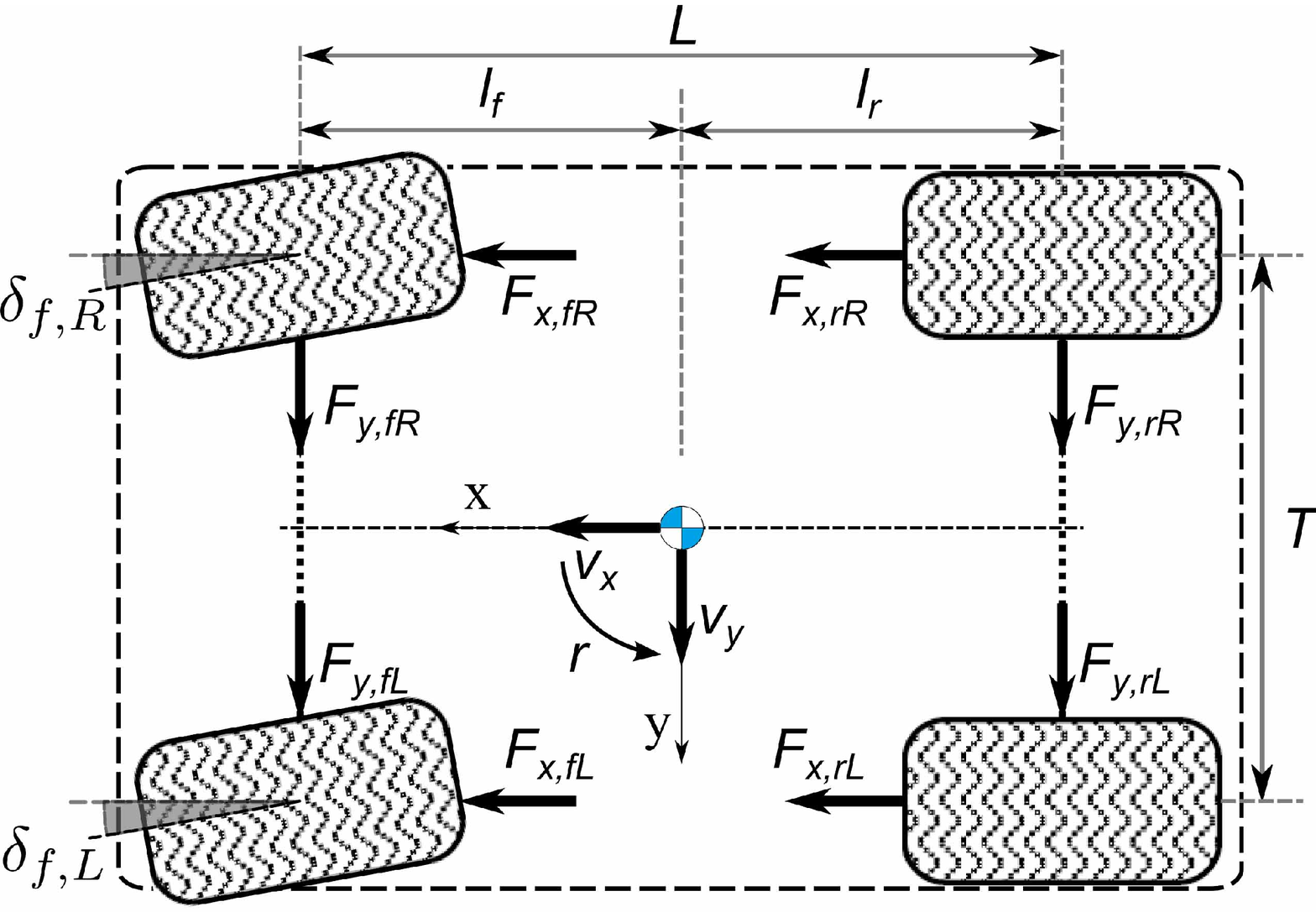}
  \label{fig:TopView}
 }
 \caption{Vehicle forces diagram.}
 \label{fig:VehicleViews}
\end{figure}

Assuming that the sprung mass rotates about the \ac{CM} of the undercarriage, that the car inertia matrix is diagonal, and that all the car wheels touch the ground, the car nonlinear model is defined by:
%------- Equation -----------------%
\begin{subequations}
 \begin{align}
  F_{x,T} &= m \left( \dot{u} - v r \right) + m_{SM} h_{SM} p \cos \phi,\\
  F_{y,T} &= m \left( \dot{v} + u r \right) - m_{SM} h_{SM} \left( \dot{p} \cos \phi - p^2 \sin \phi \right),\\
  L_T &= - K_s \left( 1 - \overline{ \Delta k }_{ss}^2 \right) \tan \phi -\notag\\
  & D_s \left( 1 - \overline{ \Delta d }_{ss}^2 \right) p \cos \phi - m g \left( \overline{ \Delta k }_{ss} + \overline{ \Delta d }_{ss} \right),\\
  N_T &= I_{zz} \dot{r},\\
  \dot{p} &= \frac{ h_{SM} m_{SM} \left( \frac{ F_{y,T} }{ m } + \sin \phi \left( g + h_{SM} \frac{ m_{UC} }{ m } p^2 \right) \right) + L_T }{ I_{xx,SM} + h^2_{SM} m_{SM} \frac{ m_{UC} }{ m } \cos \phi }.\notag
 \end{align}
 \label{eq:Dynamics_General}
\end{subequations}
%------- Equation -----------------%
Most of the model parameters are illustrated in Figure~\ref{fig:VehicleViews} and their values are given in Table~\ref{tab:Results_Vehicle_SimParam} for the simulation model used.
The model states are the vehicle velocity components in the horizontal plane ($ u, v $), roll ($ \phi $), roll rate ($ p $), and turn rate ($ r $).
$ F_{x,T} $, $ F_{y,T} $, $ L_{T} $, and $ N_{T} $ are the forces and moments acting on the car through the tires.
$ m_{SM} $, $ m_{UC} $, $ m $, $ I_{xx,SM} $ and $ I_{zz} $ are the sprung mass, the undercarriage, and the overall vehicle mass and inertia moments, respectively.
$ K_s $ and $ D_s $ are the suspension roll stiffness and damping coefficients, and $ \overline{ \Delta k }_{ss} $ and $ \overline{ \Delta d }_{ss} $ are the suspension roll differential stiffness and damping factors.

%========= Figure ===============%
%========= Figure ===============%

The simulation results (sec.~\ref{sec:Results}) include some instants where the wheels on one of the sides lift from the road.
In such conditions, the car dynamics are similar to a two segment inverted pendulum \cite{ARCLab2015CarUndercarriage}.
For the sake of readability and because it is not relevant for the design of the reference governors that maintain vehicle operation away from this condition, the extension of the vehicle equations of motion for the wheel lift condition is not presented here.

%=====================================================================================%
\subsubsection{Magic Formula Tire Model}
\label{sec:Model_NonlinearCar_MagicFormula}
%========= Figure ===============%
\begin{figure}[b]
 \centering
 \includegraphics[width = 1.5in]{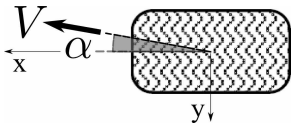}
 \caption{Tire slip angle ($ \alpha $) at the contact patch. $ V $ is the total tire speed.}
 \label{fig:TireForces_SideSlip}
\end{figure}
%========= Figure ===============%

%%========= Figure ===============%
%\begin{figure*}[!t]
% \centering
% \subfloat[Top view.]{
%  \includegraphics[height = 2.2in]{TopView}
%  \label{fig:TopView}
% }
% \hspace{1cm}
%% \centering
%% \subfloat[Side view.]{
%%  \includegraphics[height = 2.25in]{SideView}
%%  \label{fig:SideView}
%% }
% \subfloat[Rear view.]{
%  \includegraphics[height = 2.25in]{RearView}
%  \label{fig:RearView}
% }
% \caption{Vehicle forces diagram.}
% \label{fig:VehicleViews}
%\end{figure*}
%%========= Figure ===============%

The main source of nonlinearities in the equations of motion is the tire forces' dependence on slip.
In this work, we use the \emph{Magic Formula} tire model \cite{Karnopp2013CarDynamicsControl,Ulsoy2012CarDynamicsControl,Hong2013TireCoeffEstimation,Stein1980TruckModel}.
To compute the tire forces, the slip ratio ($ \lambda $) and the tire slip angle ($ \alpha $) are defined as (fig.~\ref{fig:TireForces_SideSlip}):
%------- Equation -----------------%
\begin{subequations}
 \begin{align}
  \lambda &= \begin{cases}
   \frac{ R_w \omega_w - u_w }{ u_w } & R_w \omega_w < u_w\\
   \frac{ R_w \omega_w - u_w }{ R_w \omega_w } & R_w \omega_w \geq u_w
  \end{cases},
  \label{eq:TireModels_TireSlipRatio}\\
  \alpha_f &= \delta_f - \tan^{-1} \frac{ v + l_f r }{ u },% = \tan^{-1} \frac{ u \sin \delta_f - \left( v + l_f r \right) \cos \delta_f }{ u \cos \delta_f + \left( v + l_f r \right) \sin \delta_f },
  \label{eq:TireModels_TireSlip_Front}\\
  \alpha_r &= \tan^{-1} \frac{ - v + l_r r }{ u }.
  \label{eq:TireModels_TireSlip_Rear}
\end{align}
 \label{eq:TireModels_SlipRatio&TireSlip}
\end{subequations}
%------- Equation -----------------%
%========= Figure ===============%
\begin{figure}[b]
 \centering
 \includegraphics[trim=0.5cm 0.0cm 1.1cm 0.0cm, clip=true, width = 2.5in]{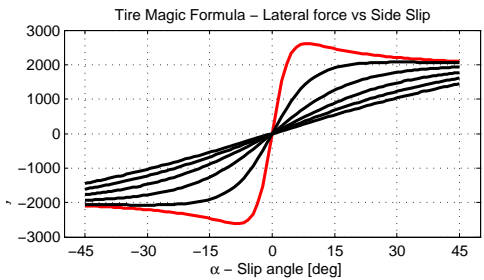}
 \caption{Tire lateral force variation with the side slip for several levels of slip ratio.}
 \label{fig:TireForces_MagicFormula_Combined}
\end{figure}
%========= Figure ===============%
The tire slip ratio ($ \lambda $) characterizes the longitudinal slip.
It is defined as the normalized difference between wheel hub speed ($ u_w $) and wheel circumferential speed ($ R_w \omega_w $).
The tire slip angle ($ \alpha $) is the angle between the tire heading direction and the velocity vector at the contact patch (fig.~\ref{fig:TireForces_SideSlip}).
Equations \eqref{eq:TireModels_TireSlip_Front} and \eqref{eq:TireModels_TireSlip_Rear} define the tire slip angle for the forward and rear wheels, respectively.
For combined longitudinal and lateral slip, the Magic Formula takes the following form \cite{Zhou2009Car_ActiveSafety_Impacts} (fig.~\ref{fig:TireForces_MagicFormula_Combined}):
%------- Equation -----------------%
\begin{align}
  \left[\begin{array}{c}
   F_x^T\\
   F_y^T
  \end{array}\right] &= F_P P \left( s_c, C, E \right) \hat{\mathbf{s}},
  \label{eq:MagicFormulaModel4CombinedSlip_MagicFormula}\\
  P \left( s_c, C, E \right) &= \sin \left( C \tan^{-1} \left[ \frac{ s_c }{ C } \left( 1 - E \right) + E \tan^{-1} \left( \frac{ s_c }{ C } \right) \right] \right),
  %\label{eq:MagicFormulaModel4CombinedSlip_NormalizedMagicFormula}
  \notag\\
  s_c &= \frac{ C_{ \alpha } \left\| \mathbf{ s } \right\| }{ F_p }, \:\:\: C_{ \alpha } = c_1 m g \left( 1 - e^{ - \frac{ c_2 F_z }{ m g } } \right),
  %\label{eq:MagicFormulaModel4CombinedSlip_CombinedSlip}
  \notag\\
  %\label{eq:MagicFormulaModel4CombinedSlip_CorneringStiffness}\\
  c_1 &= \frac{ B C D }{ 4 \left( 1 - e^{-\frac{ c_2 }{4}} \right) },
  %\label{eq:MagicFormulaModel4CombinedSlip_CorneringStiffness_c1}
  \notag\\
  F_P &= \frac{ F_z 1.0527 D }{ 1 + \left( \frac{ 1.5 F_z }{ m g } \right)^3 },
  %\label{eq:MagicFormulaModel4CombinedSlip_PeakTierForce}
  \notag\\
  \mathbf{s} &= \left[\begin{array}{c}
   s_x\\
   s_y
  \end{array}\right] = \left[\begin{array}{c}
   \lambda\\
   \tan \alpha
  \end{array}\right], \:\:\: \hat{ \mathbf{ s } } = \frac{ \mathbf{ s } }{ \left\| \mathbf{ s } \right\| },
  %\label{eq:MagicFormulaModel4CombinedSlip_SlipVector}
  \notag
  %\label{eq:MagicFormulaModel4CombinedSlip_NormalizedSlipVector}
%\label{eq:MagicFormulaModel4CombinedSlip}
\end{align}
%------- Equation -----------------%
where $ F_x^T $ and $ F_y^T $ are the forces along the tire longitudinal and lateral axis, $ x $ and $ y $ (fig.~\ref{fig:TireForces_SideSlip}), respectively. $ C_{\alpha} $ is the cornering stiffness, $ F_P $ is the horizontal (or slip) peak force, $ \hat{ \mathbf{ s } } $ is the total slip, and $ B $, $ C $, $ D $, $ E $, and $ c_2 $ are tire parameters that depend on the tire characteristics and the road conditions (see Table~\ref{tab:Model_Tires_MFParam}).
%%========= Figure ===============%
%\begin{figure}[b]
% \centering
% \includegraphics[trim=0.5cm 0.0cm 1.1cm 0.0cm, clip=true, width = 2.5in]{TireMagicFormula_LatForce_vs_SideSlip_SlipRatio}
% \caption{Tire lateral force variation with the side slip for several levels of slip ratio.}
% \label{fig:TireForces_MagicFormula_Combined}
%\end{figure}
%%========= Figure ===============%

%---------- Table ---------------%
\begin{table}[!h]
 \footnotesize
 \caption{Tires Magic Formula model parameters.}
 \label{tab:Model_Tires_MFParam}
 \begin{centering}
  \begin{tabular}{|l||c|c|c|c|c|}
   \hline Conditions & $ B $ & $ C $ & $ D $ & $ E $ & $ c_2 $ \tabularnewline
   \hline
   \hline Dry & $ 7.15 $ & $ 2.30 $ & $ 0.87 $ & $ 1.00 $ & $ 1.54 $ \tabularnewline
   \hline Wet & $ 9.00 $ & $ 2.50 $ & $ 0.72 $ & $ 1.00 $ & $ 1.54 $ \tabularnewline
   \hline Snow & $ 5.00 $ & $ 2.00 $ & $ 0.30 $ & $ 1.00 $ & $ 1.54 $ \tabularnewline
   \hline Ice & $ 4.00 $ & $ 2.00 $ & $ 0.10 $ & $ 1.00 $ & $ 1.54 $ \tabularnewline
   \hline
  \end{tabular}
  \par
 \end{centering}
\end{table}
%---------- Table ---------------%
%Figure~\ref{fig:TireForces_RoadConditions} illustrates the variation of the lateral force with the road conditions, based on the parameters in Table~\ref{tab:Model_Tires_MFParam}.
%The side slip angle at which the maximum lateral force occurs varies with the different road conditions.
%The pronounced peak with wet tarmac produces acute variations in the car dynamics, when the tires transition from the almost linear positive slope before the peak to the pronounced negative slope beyond the peak.
%Note also that the initial slope inclination is identical for dry and wet tarmac, and very shallow for snow and ice.
%%========= Figure ===============%
%\begin{figure}[t]
% \centering
% \includegraphics[trim=0.5cm 0.0cm 1.1cm 0.0cm, clip=true, width = 2.5in]{TireMagicFormula_2D_LatForce_vs_SideSlip_RoadCond}
% \caption{Tire lateral force variation with the road conditions. The red solid line represents the lateral force for dry tarmac, while the other three, in decreasing order of magnitude, correspond to wet tarmac, snow, and ice conditions.}
% \label{fig:TireForces_RoadConditions}
%\end{figure}
%%========= Figure ===============%

\begin{figure}[t]
 \centering
 \includegraphics[trim=0.5cm 0.0cm 1.1cm 0.0cm, clip=true, width = 2.5in]{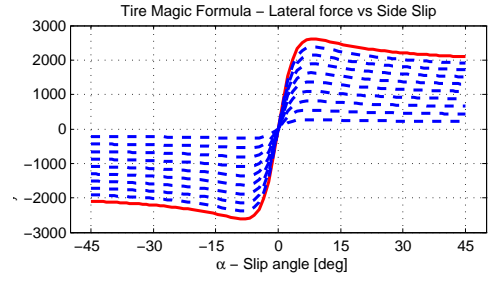}
 \caption{Tire lateral force variation with the vertical load. The solid red line is the lateral force generated at nominal vertical load.}
 \label{fig:TireForces_VertLoad}
\end{figure}

Figure~\ref{fig:TireForces_VertLoad} illustrates the variation of the lateral force with the vertical load.
The illustrated model presents almost the same side slip angle for all vertical load cases, and an initial slope decreasing for lower vertical loads.
This behavior is governed by the parameter $ c_2 $.
%========= Figure ===============%
%========= Figure ===============%

%% file: Model_Constraints.tex
\subsection{Constraints}
\label{sec:Model_Constraints}

The vehicle constraints reflect the system's physical limits or requirements to keep the vehicle state in a safe region.
There are several options to define constraints that protect the vehicle from rollover.
One possibility is to define the rollover constraints as the states where the car roll is so large that there is no possible recovery.
This approach would require the treatment of complex, hybrid vehicle dynamics, which involve the vehicle continuous dynamics states plus two discrete states (all wheels on the road state and wheel liftoff state), and in the case of wheel liftoff, multiple-body inverted pendulum-like dynamics.

In this paper, following also \cite{Solmaz2006RolloverPrevention_DifferentialBraking,Solmaz2007RolloverPrevention_ActiveSteering_Conf,Solmaz2007RolloverPrevention_ActiveSteering}, more conservative rollover avoidance constraints are treated, which are, however, simpler to enforce.
These constraints are defined through the \acf{LTR}.
The \ac{LTR} measures how much of the vehicle vertical load is concentrated on one of the vehicle sides:
%------- Equation -----------------%
\begin{equation}
 LTR := \frac{ F_{z,R} - F_{z,L} }{ m g }.
 \label{eq:Constr_LTR_def}
\end{equation}
%------- Equation -----------------%

The wheel liftoff happens when the \ac{LTR} increases above $ 1 $ or decreases below $ -1 $, \emph{i.e.}, when the right or left wheels, respectively, bear all the car's weight.
Hence, the rollover avoidance constraints are imposed as:
%------- Equation -----------------%
\begin{equation}
 LTR \in \left[ - LTR_{lim}, LTR_{lim} \right], \:\: 0 < LTR_{lim} < 1.
 \label{eq:Constr_LTR_constr}
\end{equation}
%------- Equation -----------------%

\begin{remark}
 Note that the absolute value of the \ac{LTR} may exceed 1, even if wheel liftoff does not occur.
 This can happen, in particular, due to the suspension roll moment, \emph{e.g.}, generated by the spring stored energy, and by the \ac{CM} vertical acceleration during wheel liftoff.
\end{remark}

The steering input is also considered to be constrained:
%All the control variables are also considered to be constrained:
%------- Equation -----------------%
\begin{equation}
 \delta_{SW} \in \left[ - \delta_{SW,lim}, \delta_{SW,lim} \right].%,\\
 \label{eq:Constr_Ctrl_constr}
\end{equation}
%------- Equation -----------------%
%%------- Equation -----------------%
%\begin{subequations}
% \begin{align}
% \delta_{SW} &\in \left[ - \delta_{SW,lim}, \delta_{SW,lim} \right],\\
% \delta_{DB} &\in \left[ - \delta_{DB,lim}, \delta_{DB,lim} \right],\\
% \Delta K_{ss} &\in \left[ - \Delta K_{ss,lim}, \Delta K_{ss,lim} \right],\\
% \Delta D_{ss} &\in \left[ - \Delta D_{ss,lim}, \Delta D_{ss,lim} \right],\\
% \overline{ \Delta k }_{ss} &\in \left[ - \overline{ \Delta k }_{ss,lim}, \overline{ \Delta k }_{ss,lim} \right],\\
% \overline{ \Delta d }_{ss} &\in \left[ - \overline{ \Delta d }_{ss,lim}, \overline{ \Delta d }_{ss,lim} \right].
% \end{align}
% \label{eq:Constr_Ctrl_constr}
%\end{subequations}
%%------- Equation -----------------%

%% file: Model_LinearCar.tex
\subsection{Linearized Car Model}
\label{sec:Model_LinearCar}

The vehicle linear model has the following form,
%%------- Equation -----------------%
%\begin{subequations}
% \begin{align}
%  \dot{\mathbf{ x }} &= \mathbf{ A } \mathbf{ x } + \mathbf{B}_{ \delta_S } k_{\delta_{SW}} \delta_{SW} + \mathbf{B}_{ \delta_{DB} } \delta_{DB} + \mathbf{B}_{ \Delta K_{ss} } \Delta K_{ss} +\notag\\
%  &\mathbf{B}_{ \Delta D_{ss} } \Delta D_{ss} + \mathbf{B}_{ \overline{ \Delta k }_{ss} } \overline{ \Delta k }_{ss} + \mathbf{B}_{ \overline{ \Delta d }_{ss} } \overline{ \Delta d }_{ss},\\
%  \mathbf{ y } &= \mathbf{ C } \mathbf{ x } + \mathbf{D}_{ \delta_{SW} } \delta_{SW} + \mathbf{D}_{ \delta_{DB} } \delta_{DB} + \mathbf{D}_{ \Delta K_{ss} } \Delta K_{ss} +\notag\\
%  &\mathbf{D}_{ \Delta D_{ss} } \Delta D_{ss} + \mathbf{D}_{ \overline{ \Delta k }_{ss} } \overline{ \Delta k }_{ss} + \mathbf{D}_{ \overline{ \Delta d }_{ss} } \overline{ \Delta d }_{ss},
% \end{align}
% \label{eq:LinModel_Dynamics}
%\end{subequations}
%%------- Equation -----------------%
%------- Equation -----------------%
\begin{subequations}
 \begin{align}
  \dot{\mathbf{ x }} &= \mathbf{ A } \mathbf{ x } + \mathbf{B}_{ \delta_S } k_{\delta_{SW}} \Delta \delta_{SW},\\
  \mathbf{ y } &= \mathbf{ C } \mathbf{ x } + \mathbf{D}_{ \delta_{SW} } \Delta \delta_{SW},
 \end{align}
 \label{eq:LinModel_Dynamics}
\end{subequations}
%------- Equation -----------------%
\hspace{-0.2cm}
where $ \mathbf{ x } = \left[ \Delta v, \Delta p, \Delta r, \Delta \phi \right] $ correspond to the relevant lateral state variables.
$ \Delta \delta_{SW} $ is the steering control input deviation.
The ratio between the steering wheel angle and the forward tires steering angles is $ k_{\delta_{SW}} $.

Linearizing \eqref{eq:Dynamics_General}, the vehicle linearized model is obtained, with the dynamics matrix of the form,
%\textbf{ToDo: \emph{Add the latest symbolic linearization to compare in terms of value and readability.}}
%\textbf{ToDo: \emph{Reference to numerical linearization.}}
%------- Equation -----------------%
\begin{equation}
 \mathbf{A} := \left[ \begin{array}{cccc}
  \frac{ \partial F_{y,T} }{ \partial v } \frac{ 1 }{ m' } & \frac{ \partial F_{y,T} }{ \partial r } \frac{ 1 }{ m' } - u & h'_{SM} \frac{ \partial \dot{ p } }{ \partial p } & h'_{SM} \frac{ \partial \dot{ p } }{ \partial \phi } \\
  \frac{ \partial N_{T} }{ \partial v } \frac{ 1 }{ I_{ zz } } & \frac{ \partial N_{T} }{ \partial r } \frac{ 1 }{ I_{ zz } } & 0 & 0 \\
  \frac{ \partial F_{y,T} }{ \partial v } \frac{ h_{SM} }{ I'_{ xx } } & \frac{ \partial F_{y,T} }{ \partial r } \frac{ h_{SM} }{ I'_{ xx } } & \frac{ \partial \dot{ p } }{ \partial p } & \frac{ \partial \dot{ p } }{ \partial \phi } \\
  0 & 0 & 1 & 0 
 \end{array} \right],
 \label{eq:LinModel_DynamicsMatrix}
\end{equation}
%------- Equation -----------------%
where
%------- Equation -----------------%
\begin{subequations}
 \begin{align}
  m' &= \frac{ m^2 I'_{ xx } }{ m I'_{ xx } + h^2_{SM} m^2_{SM} \cos \phi_0 },\\
  I'_{ xx } &= I_{ xx,SM } + h^2_{SM} m_{SM} \frac{ m_{UC} }{ m } \cos \phi,\\
  h'_{SM} &= \frac{ h_{SM} m_{SM} }{ m },\\
  \frac{ \partial \dot{ p } }{ \partial p } &= \frac{ 2 h^2_{SM} m_{SM} p_0 \frac{ m_{UC} }{ m } \sin \phi_0 - D_s \left( 1 - \overline{ \Delta d }_{ss,0}^2 \right) \cos \phi_0 }{ I'_{ xx } },\\
  \frac{ \partial \dot{ p } }{ \partial \phi } &= \left[ m_{SM} g h_{SM} - K_s \left( 1 - \overline{ \Delta k }_{ss,0}^2 \right) \left( 1 + \tan^2 \phi_0 \right) - \right.\notag\\
  &\left. D_s \left( 1 - \overline{ \Delta d }_{ss,0}^2 \right) p_0 \sin \phi_0 \right] / I'_{ xx }.
 \end{align}
 \label{eq:LinModel_AuxVar}
\end{subequations}
%------- Equation -----------------%

The steering control matrix is defined as:
%------- Equation -----------------%
\begin{equation}
 \mathbf{B}_{ \delta_S } := \left[ \begin{array}{cccc}
  \frac{ \partial F_{y,T} }{ \partial \delta_f } \frac{ 1 }{ m' } &
  \frac{ \partial N_{T} }{ \partial \delta_f } \frac{ 1 }{ I_{ zz } } &
  \frac{ \partial F_{y,T} }{ \partial \delta_f } \frac{ h_{SM} }{ I'_{ xx } } &
  0 
 \end{array} \right]^\intercal,
 \label{eq:LinModel_ControlMatrix_Steer}
\end{equation}
%------- Equation -----------------%
%while the differential braking control matrix is:
%%------- Equation -----------------%
%\begin{equation}
% \mathbf{B}_{ \delta_{DB} } := \left[ \begin{array}{cccc}
%  0 &
%  \frac{ T }{ 2 I_{ zz } } &
%  0 &
%  0 
% \end{array} \right]^\intercal,
% \label{eq:LinModel_ControlMatrix_DiffBrak}
%\end{equation}
%%------- Equation -----------------%
%and the suspension control matrix is:
%%------- Equation -----------------%
%\begin{subequations}
% \begin{align}
%  \mathbf{B}_{ \Delta K_{ss} } &:= \left[ \begin{array}{cccc}
%   0 &
%   0 &
%   - \frac{ \left( 1 - \overline{ \Delta k }_{ss,0}^2 \right) \tan \phi_0 }{ I'_{ xx } } &
%   0 
%  \end{array} \right]^\intercal,\\
%  \mathbf{B}_{ \Delta D_{ss} } &:= \left[ \begin{array}{cccc}
%   0 &
%   0 &
%   - \frac{ \left( 1 - \overline{ \Delta d }_{ss,0}^2 \right) p_0 \cos \phi_0 }{ I'_{ xx } } &
%   0 
%  \end{array} \right]^\intercal,\\
%  \mathbf{B}_{ \overline{ \Delta k }_{ss} } &:= \left[ \begin{array}{cccc}
%   0 &
%   0 &
%   \frac{ K_s 2 \overline{ \Delta k }_{ss,0} \tan \phi_0 - m g  }{ I'_{ xx } } &
%   0 
%  \end{array} \right]^\intercal,\\
%  \mathbf{B}_{ \overline{ \Delta d }_{ss} } &:= \left[ \begin{array}{cccc}
%   0 &
%   0 &
%   \frac{ D_s 2 \overline{ \Delta d }_{ss,0} p_0 \cos \phi_0 - m g  }{ I'_{ xx } } &
%   0 
%  \end{array} \right]^\intercal.
% \end{align}
% \label{eq:LinModel_ControlMatrices_Susp}
%\end{subequations}
%%------- Equation -----------------%

The system output matrices are defined by the operation constraints (sec.~\ref{sec:Model_Constraints}):
%------- Equation -----------------%
\begin{subequations}
 \begin{align}
  \mathbf{C} &:= \left[ \begin{array}{cccc}
   0 & 0 &  \frac{2 D_{s} }{ m g T} & \frac{2 K_{s} }{ m g T} \\
%   0 & 0 & 0 & 0 \\
%   0 & 0 & 0 & 0 \\
%   0 & 0 & 0 & 0 \\
%   0 & 0 & 0 & 0 \\
%   0 & 0 & 0 & 0 \\
   0 & 0 & 0 & 0 
  \end{array} \right],
  \label{eq:LinModel_OutputMatrices_C} \\
  \mathbf{D}_{ \delta_S } &:= \left[ \begin{array}{ccccccc}
   0 & 1 %& 0 & 0 & 0 & 0 & 0
%  \end{array} \right]^\intercal, \\
%  \mathbf{D}_{ \delta_{DB} } &:= \left[ \begin{array}{ccccccc}
%   0 & 0 & 1 & 0 & 0 & 0 & 0
%  \end{array} \right]^\intercal, \\
%  \mathbf{D}_{ \Delta K_{ss} } &:= \left[ \begin{array}{ccccccc}
%   0 & 0 & 0 & 1 & 0 & 0 & 0
%  \end{array} \right]^\intercal,\\
%  \mathbf{D}_{ \Delta D_{ss} } &:= \left[ \begin{array}{ccccccc}
%   0 & 0 & 0 & 0 & 1 & 0 & 0
%  \end{array} \right]^\intercal,\\
%  \mathbf{D}_{ \overline{ \Delta k }_{ss} } &:= \left[ \begin{array}{ccccccc}
%   0 & 0 & 0 & 0 & 0 & 1 & 0
%  \end{array} \right]^\intercal,\\
%  \mathbf{D}_{ \overline{ \Delta d }_{ss} } &:= \left[ \begin{array}{ccccccc}
%   0 & 0 & 0 & 0 & 0 & 0 & 1
  \end{array} \right]^\intercal.
 \end{align}
 \label{eq:LinModel_OutputMatrices}
\end{subequations}
%------- Equation -----------------%

%% file: Model_PerformanceMetrics.tex
\subsection{Performance Metrics}
\label{sec:Model_PerformanceMetrics}

This study uses four performance metrics: the step computation time, the effectiveness, the conservatism, and the turning response.
We have chosen not to use a metric that compares the vehicle positions between the reference trajectory and the trajectory affected by the controllers as there are several reference trajectories that end in a full rollover.

The step computation time is the time it takes the controller to perform the constraint enforcement verification and compute a command.
%The effectiveness metric is the success rate in avoiding rollover:
%%------- Equation -----------------%
%\begin{equation}
% \eta := \frac{N - N_{rollover}}{N},
% \label{eq:Model_Efficiency}
%\end{equation}
%%------- Equation -----------------%
%where $ N $ is the total number of tests and $ N_{rollover} $ is the number of tests where the car rolls-over.
The effectiveness metric is the success rate in avoiding wheel lift up to a wheel lift limit.
For each test:
%------- Equation -----------------%
\begin{equation}
 \eta_{lift} := 1 - \frac{\max z_{wheel}}{z_{wheel,lim}},
% \eta := 1 - \frac{\max_{t \in \left[ 0, T \right] } z_{wheel}}{z_{wheel,lim}},
 \label{eq:Model_Efficiency}
\end{equation}
%------- Equation -----------------%
where $ \max z_{wheel} $ is the maximum wheel lift attained during a test, and $ z_{wheel,lim} $ is the wheel lift limit considered.

The conservatism metric indicates how much the controller is over-constraining the steering command when compared with a safe steering command:
%The conservatism metric indicates how much the controller command differs from the reference steering command compared:% when no rollover would occur even without controller activation:
%%------- Equation -----------------%
%\begin{equation}
% \chi := \frac{\delta_{SW,ref,peak} - \delta_{SW,peak}}{\delta_{SW,ref,peak}},
% \label{eq:Model_Conservatism}
%\end{equation}
%%------- Equation -----------------%
%where $ \delta_{SW,ref,peak} $ is the reference command peak value during the maneuver and $ \delta_{SW,peak} $ is the peak command allowed by the controller.
%------- Equation -----------------%
\begin{equation}
 \chi := \frac{ \int_0^T \left( \left| \delta_{SW,ref} - \delta_{SW} \right| - \left| \delta_{SW,ref} - \delta_{SW,safe} \right| \right) d t }{ \int_0^T \left| \delta_{SW,ref} \right| d t },
 \label{eq:Model_Conservatism}
\end{equation}
%------- Equation -----------------%
%{\small
%%------- Equation -----------------%
%\begin{equation}
% \chi := \frac{ \int_0^T \left( \left| \delta_{SW,ref} \left( t \right) - \delta_{SW} \left( t \right) \right| - \left| \delta_{SW,ref} \left( t \right) - \delta_{SW,safe} \left( t \right) \right| \right) d t }{ \int_0^T \left| \delta_{SW,ref} \left( t \right) \right| d t },
% \label{eq:Model_Conservatism}
%\end{equation}
%%------- Equation -----------------%
%}
where $ \delta_{SW,ref} $ is the driver reference command during the maneuver, $ \delta_{SW} $ is the controller command, $ \delta_{SW,safe} $ is a reference safe command, and $ T $ is the test duration.
Two options for the reference safe command are the driver steering input scaled down to avoid any wheel lift or the driver steering input scaled down to produce a maximum wheel lift of $ z_{wheel,lim} $.
%\textbf{Should the conservatism also take into account the other reference commands? \emph{Differencial Braking and Active Suspension?} A metric centered on the steering command conservatism may skew the results.}
%\textbf{ToDo: \emph{Complement the conservatism definition with the relative conservatism.}}
%%------- Equation -----------------%
%\begin{equation}
% \chi_{rel} := \chi_{ctrl} - \chi_{ref},
% \label{eq:Model_Conservatism_Rel}
%\end{equation}
%%------- Equation -----------------%

The turning response metric indicates how much the controller is limiting the vehicle turn rate relative to the driver desired turning rate when compared with the turn rate achieved with a safe steering command:
%------- Equation -----------------%
\begin{equation}
 \eta_{\psi} := \frac{ \int_0^T \left( \left| \psi_{des} - \psi_{safe} \right| - \left| \psi_{des} - \psi \right| \right) d t }{ \int_0^T \left| \psi_{SW,ref} \right| d t },
 \label{eq:Model_TurnResp}
\end{equation}
%------- Equation -----------------%
%%------- Equation -----------------%
%\begin{equation}
% \eta_{\psi} := \frac{ \int_0^T \left( \left| \psi_{des} \left( t \right) - \psi_{safe} \left( t \right) \right| - \left| \psi_{des} \left( t \right) - \psi \left( t \right) \right| \right) d t }{ \int_0^T \left| \psi_{SW,ref} \left( t \right) \right| d t },
% \label{eq:Model_TurnResp}
%\end{equation}
%%------- Equation -----------------%
where the driver desired turning rate is inferred from the reference steering command and the steering-to-turn rate stability derivative: $ \psi_{des} = \left. \frac{d\psi}{d\delta_{SW}} \right|_{\delta_{SW} = 0} \delta_{SW,ref} $.
$ \psi $ and $ \psi_{safe} $ are the turn rates caused by the controller command and the reference safe command, respectively.

%% file: Ctrl_RG.tex
\section{Reference and Command Governors}
\label{sec:Ctrl_RG}

\begin{figure}[t]
 \centering
 \includegraphics[width = 2.0in]{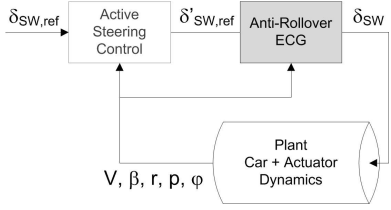}
 \caption{Command governor control architecture for a steering command.}
 \label{fig:Ctrl_RG_Arch}
\end{figure}

This work implements different versions of Reference and \acf{CG} controllers, collectively referred to by their common name as reference governors.
Our reference governors modify the reference command, the steering angle (fig~\ref{fig:Ctrl_RG_Arch}), if this reference command is predicted to induce a violation of the \ac{LTR} constraints (sec.~\ref{sec:Model_Constraints}).

The reference governors solutions studied for this application are based on both linear and nonlinear models.
The following sections describe the various solutions in more detail.

%========= Figure ===============%
%========= Figure ===============%

%=====================================================================================%
\subsection{\aclp{LRG} and \aclp{CG}}
\label{sec:Ctrl_LinearRG&CG}

Both the \aclp{LRG} and the \aclp{CG} rely on a maximum output admissible set $ \mathcal{O}_{ \infty } $ (or its subsets) to check if a reference command is safe, \emph{i.e.}, if it does not lead to constraint violation, and to compute a safe alternative command, if necessary.
The $ \mathcal{O}_{ \infty } $ set characterizes the combinations of constant commands and vehicle states that do not lead to constraint violating responses,
%------- Equation -----------------%
\begin{equation}
 \mathcal{O}_{ \infty } := \left\{ \left( \mathbf{u}, \mathbf{x} \right) | \mathbf{y}_k \in \mathbf{Y}, \forall k \in \mathbb{Z}^{0+} \right\} \subset \mathbb{R}^{4+n},
 \label{eq:Ctrl_RG_Oinf}
\end{equation}
%------- Equation -----------------%
where $ n = 1 $ is the number of control variables, $ \mathbf{x} $ and $ \mathbf{u} $ are the state and command at the present moment and $ \mathbf{y}_k, k \in \mathbb{Z}^{0+} $, is the predicted evolution of the system output.
The set $ \mathbf{Y} $ represents the constraints and delineates safe outputs
%------- Equation -----------------%
\begin{equation}
 \mathbf{Y} := \left\{ \mathbf{y} | \mathbf{A}_y \mathbf{y} \leq \mathbf{b}_y \right\} \subset \mathbb{R}^{l},
 \label{eq:Ctrl_SafeOutput}
\end{equation}
%------- Equation -----------------%
where $ l $ is the number of system outputs on which constraints are imposed.

Considering \eqref{eq:Ctrl_RG_Oinf}, \eqref{eq:Ctrl_SafeOutput}, $ \mathbf{x}_{\mathcal{O}_{ \infty }} := \left[ \begin{array}{ccc} \mathbf{u} & \mathbf{x} \end{array} \right]^{\intercal} $, and a linear model \eqref{eq:LinModel_Dynamics}, we define an inner approximation of the maximum output admissible set as
{\small
%------- Equation -----------------%
\begin{subequations}
 \begin{align}
  \mathcal{O}_{ \infty } &:= \left\{ \mathbf{x}_{\mathcal{O}_{ \infty }} | \mathbf{A}_{\mathcal{O}_{ \infty }} \mathbf{x}_{\mathcal{O}_{ \infty }} \leq \mathbf{b}_{\mathcal{O}_{ \infty }} \right\} \subset \mathbb{R}^{4+n},\\
  \mathbf{A}_{\mathcal{O}_{ \infty }} &:= \left[ \begin{array}{cc}
   \mathbf{A}_y \mathbf{D} & \mathbf{A}_y \mathbf{C} \\
   \mathbf{A}_y \left( \mathbf{C} \left( \mathbf{I} - \mathbf{A} \right)^{-1} \left( \mathbf{I} - \mathbf{A} \right) \mathbf{B} + \mathbf{D} \right) & \mathbf{A}_y \mathbf{C} \mathbf{A} \\
   \vdots & \vdots \\
   \mathbf{A}_y \left( \mathbf{C} \left( \mathbf{I} - \mathbf{A} \right)^{-1} \left( \mathbf{I} - \mathbf{A}^{N} \right) \mathbf{B} + \mathbf{D} \right) & \mathbf{A}_y \mathbf{C} \mathbf{A}^{N} \\
   \mathbf{A}_y \mathbf{H} & 0
  \end{array} \right], \\
  \mathbf{b}_{\mathcal{O}_{ \infty }} &:= \left[ \begin{array}{c}
   \mathbf{b}_y \\
   \mathbf{b}_y \\
   \vdots \\
   \mathbf{b}_y \\
   \mathbf{b}_y \left( 1 - \epsilon \right)
  \end{array} \right],\\
  H &= \mathbf{C} \left( \mathbf{I} - \mathbf{A} \right)^{-1} \mathbf{B} + \mathbf{D},
 \end{align}
 \label{eq:Ctrl_RG_OinfMatrices}
\end{subequations}
%------- Equation -----------------%
}
\hspace{-0.375cm}
where $ N $ is the selected prediction horizon and $ \epsilon > 0$.
Under mild assumptions \cite{Gilbert1995ReferenceGovernors}, $ \mathcal{O}_\infty $ is the same for all $ N $ sufficiently large and is positively invariant (for constant $ u $) and satisfies constraints pointwise.
Generally, such an N is comparable to the settling time of the system.

%=====================================================================================%
\input{Ctrl_LRG}

%=====================================================================================%

\input{Ctrl_ECG}

%=====================================================================================%
\input{Ctrl_Steering}

%=====================================================================================%
\input{Ctrl_NonlinearComp}

%=====================================================================================%
\input{Ctrl_Feasibility}

%=====================================================================================%
\input{Ctrl_NRG}

%%=====================================================================================%
%\input{Ctrl_DifferentialBraking}
%
%
%%=====================================================================================%
%\input{Ctrl_ActiveSuspension}
%
%
%%=====================================================================================%
%\input{Ctrl_CombinedControl}

%% file: Ctrl_LRG.tex
%=====================================================================================%
\subsection{\acf{LRG}}
\label{sec:Ctrl_LinearRG}

The \ac{LRG} computes a single command value on every update using the above $ \mathcal{O}_{ \infty } $ set which we re-write as %(fig.~\ref{fig:Ctrl_ECG}).
%As such, the $ \mathcal{O}_{ \infty } $ set is defined as:
%------- Equation -----------------%
\begin{align}
 \mathcal{O}_{ \infty } &:= \left\{ \left( \mathbf{u}, \mathbf{x} \right) | \mathbf{y}_k = \left( \mathbf{C}  \left( \mathbf{I} - \mathbf{A} \right)^{- 1 } \left( \mathbf{I} - \mathbf{A}^k \right) \mathbf{B} + \mathbf{D} \right) \mathbf{u} +\right.\label{eq:Ctrl_LRG_Oinf}\notag\\
 &\left. \mathbf{C} \mathbf{A}^k \mathbf{x} \in \mathbf{Y}, k = 0, ..., N \right\} \bigcap \Gamma \subset \mathbb{R}^{4+n},\\
 \Gamma &= \left\{ \left( \mathbf{u}, \mathbf{x} \right) | \mathbf{H} \mathbf{u} \in \left( 1 - \epsilon \right) \mathbf{Y} \right\}\notag.
\end{align}
%------- Equation -----------------%
By applying the $ \mathcal{O}_{ \infty } $ to the current state, the controller checks if the reference or an alternative command are safe.
If the reference command is deemed safe, it is passed to the actuator.
If not, the controller selects an alternative safe command that minimizes the difference to the reference:
%------- Equation -----------------%
\begin{align}
 \mathbf{v}_k &= \argmax_{ \mathbf{v}_k } \left\{ k_{RG} | \mathbf{v}_k = \mathbf{v}_{k - 1} + k_{RG} \left( \mathbf{u}_{k} - \mathbf{v}_{k - 1} \right), \right. \notag\\
  &\left. \left( \mathbf{v}_k, \mathbf{x}_k \right) \in \mathcal{O}_{ \infty } \: \mathsf{and} \: k_{RG} \in \left[ 0, 1 \right] \right\},
 \label{eq:Ctrl_LRG_Cmd}
\end{align}
%------- Equation -----------------%
where $ \mathbf{u}_{k} $ is the current reference command, $ \mathbf{x}_k $ is the current state, and $ \mathbf{v}_{k - 1} $ is the previous command used.

\begin{remark}
 Because the reference governor checks at each update if a command is safe for the future steps, $ \mathbf{v}_{k - 1} $ is assured to be safe for the current step, provided an accurate model of the system is used.
 As such, in the optimization process \eqref{eq:Ctrl_LRG_Cmd}, one only needs to analyze the interval $ \mathbf{u} \in \left[ \mathbf{v}_{k - 1}, \mathbf{u}_{k} \right] $.
\end{remark}

%\begin{enumerate}
% \item Command Governor activation logic
% 
% \item Command dynamics (when the Command Governor is active)
% 
% \item Constraint admissible sets
% 
% \item Command selection method
%\end{enumerate}

%% file: Ctrl_ECG.tex
%=====================================================================================%
\subsection{\acf{ECG}}
\label{sec:Ctrl_LinearECG}

The \ac{ECG} \cite{Gilbert2011ReferenceGovernors} is similar to the \ac{LRG}, but, when it detects an unsafe reference command, it computes a sequence of safe commands governed by:
%------- Equation -----------------%
\begin{subequations}
 \begin{align}
  \mathbf{v} &= \bar{\mathbf{C}} \bar{\mathbf{x}} + \mathbf{\rho},
  \label{eq:Ctrl_ECG_CommandOutput} \\
  \bar{\mathbf{x}}_{k + 1} &= \bar{\mathbf{A}} \bar{\mathbf{x}},
  \label{eq:Ctrl_ECG_CommandDynamics}
 \end{align}
 \label{eq:Ctrl_ECG_Command}
\end{subequations}
%------- Equation -----------------%
\hspace{-0.375cm}
where $ \mathbf{v} $ is the safe command (output of the \ac{ECG}) with dynamics governed by \eqref{eq:Ctrl_ECG_CommandDynamics}, $ \bar{\mathbf{x}} $ is the virtual state vector of the command dynamics, and $ \mathbf{\rho} $ is the steady state command to which the sequence converges.
The matrices $ \bar{\mathbf{A}} $ and $ \bar{\mathbf{C}} $ are two of the design elements of the \ac{ECG}.
They are defined at the end of this section. The key requirement is that the matrix $ \bar{\mathbf{A}} $ must be Schur.

To detect if a reference command is safe, the \ac{ECG} uses the \ac{LRG} $ \mathcal{O}_{ \infty } $ set \eqref{eq:Ctrl_LRG_Oinf}.
If the reference command is unsafe, the \ac{ECG} uses an augmented $ \check{\mathcal{O}}_{ \infty } $ set that takes into account the computed command dynamics:
%------- Equation -----------------%
\begin{align}
 \check{\mathcal{O}}_{ \infty } &:= \left\{ \left( \mathbf{\rho}, \mathbf{x}_{aug} \right) | \mathbf{y}_k = \right.\notag\\
 &\left. \left( \mathbf{C}_{aug} \left( \mathbf{I} - \mathbf{A}_{aug} \right)^{- 1 } \left( \mathbf{I} - \mathbf{A}_{aug}^k \right) \mathbf{B}_{aug} + \mathbf{D}_{aug} \right) \mathbf{\rho} + \right.\notag\\
 &\left. \mathbf{C}_{aug} \mathbf{A}_{aug}^k \mathbf{x}_{aug} \in \mathbf{Y}, k = 0, ..., N \right\}\bigcap \Gamma' \subset \mathbb{R}^{4+n+m},
 \label{eq:Ctrl_LinearECG_Oinf}\\
 \Gamma' &= \left\{ \left( \mathbf{\rho}, \mathbf{x}_{aug} \right) | \mathbf{H} \mathbf{\rho} \in \left( 1 - \epsilon \right) \mathbf{Y} \right\}\notag,
\end{align}
%------- Equation -----------------%
where $ m $ is the size of $ \bar{\mathbf{x}} $, $ \mathbf{x}_{aug}^{\intercal} := \left[ \mathbf{x}^{\intercal}, \bar{\mathbf{x}}^{\intercal} \right]^{\intercal} $ is an augmented state vector, and the matrices $ A_{aug}, B_{aug}, C_{aug}, D_{aug} $ correspond to the augmented system.
We note that this event-triggered execution of \ac{ECG} is quite effective in reducing average chronometric loading and to the authors' knowledge, has not been previously proposed elsewhere.

%In the \ac{LRG} only a constant command needs to be optimized.
In the \ac{ECG}, both the steady state command and the initial virtual state are optimized.
This optimization is performed by solving the following quadratic programming problem:
%------- Equation -----------------%
%\begin{subequations}
 \begin{align}
  \mathbf{x}' &= \argmin_{ \mathbf{x}' } \left\{ \frac{1}{2} \mathbf{x}'^{\intercal} \mathbf{H} \mathbf{x}' + \mathbf{f} \mathbf{\rho} | \left( \mathbf{\rho}, \mathbf{x}_{aug} \right) \in \check{\mathcal{O}}_{ \infty } \right\},
  \label{eq:ECG_QP_CostMin} \\
  \mathbf{x}' &:= \left[ \begin{array}{c}
   \bar{\mathbf{x}} \\
   \rho
  \end{array} \right], \: \mathbf{H} := \left[ \begin{array}{cc}
   \mathbf{P} & 0 \\
   0 & \mathbf{Q}
  \end{array} \right], \: \mathbf{f} := - \mathbf{u}_k^{\intercal} \mathbf{Q}, \notag
 \end{align}
%\end{subequations}
%------- Equation -----------------%
where $ \mathbf{P} $ and $ \mathbf{Q} $ are symmetric positive-definite matrices. % defined offline.
In this work $ \mathbf{Q} := k_L \mathbf{I} > 0 $ is the tuning matrix, while $ \mathbf{P} $ is computed by solving the discrete-time Lyapunov equation:
%------- Equation -----------------%
\begin{equation}
 \bar{\mathbf{A}}^{\intercal} \mathbf{P} \bar{\mathbf{A}} - \mathbf{P} + \mathbf{Q} = 0.
 \label{eq:LyaponovEquation_DiscreteTime}
\end{equation}
%------- Equation -----------------%

The safe command is then computed by \eqref{eq:Ctrl_ECG_CommandOutput}.
%%========= Figure ===============%
%\begin{figure}[t]
% \centering
% \subfloat[Steering command.]{
%  \includegraphics[height = 2.0in]{ECG_Cmd}
%  \label{fig:Ctrl_ECG_Command}
% }
% \\
% %\hfill
% \subfloat[Vehicle lateral response.]{
%  \includegraphics[height = 2.0in]{ECG_Ctrl}
%  \label{fig:Ctrl_ECG_Dynamics}
% }
% \caption{Constraint enforcement control: \ac{LRG} and \ac{ECG} versus the driver reference command (Ref).}
% \label{fig:Ctrl_ECG}
%\end{figure}
%%========= Figure ===============%

%Reference \cite{Kolmanovsky2014CommandGovernor} suggests several possibilities for the definitions of the matrices $ \bar{\mathbf{A}} $ and $ \bar{\mathbf{C}} $.
Several possibilities for the matrices $ \bar{\mathbf{A}} $ and $ \bar{\mathbf{C}} $ exist \cite{Kolmanovsky2014CommandGovernor}.
These matrices can define a shift register sequence, making the \ac{ECG} behave like a \ac{MPC}, or a Laguerre sequence, as follows:
%------- Equation -----------------%
\begin{subequations}
 \begin{align}
  \bar{\mathbf{A}} &= \left[ \begin{array}{ccccc}
   \alpha \mathbf{I}_m & \mu \mathbf{I}_m & - \alpha \mu \mathbf{I}_m & \hdots & \left( - \alpha \right)^{N-2} \mu \mathbf{I}_m \\
   0 & \alpha \mathbf{I}_m & \mu \mathbf{I}_m & \hdots & \left( - \alpha \right)^{N-3} \mu \mathbf{I}_m \\
   0 & 0 & \alpha \mathbf{I}_m & \ddots & \vdots \\
   0 & 0 & 0 & \hdots & \mu \mathbf{I}_m \\
   0 & 0 & 0 & \hdots & \alpha \mathbf{I}_m
  \end{array} \right],
  \label{eq:Ctrl_Steering_ECG_CommandDynamics} \\
  \bar{\mathbf{C}} &= \left[ \begin{array}{ccccc} 
   \mathbf{I}_m & - \alpha \mathbf{I}_m & \alpha^2 \mathbf{I}_m & \hdots & \left( - \alpha \right)^{N-1} \mathbf{I}_m
  \end{array} \right],
  \label{eq:Ctrl_Steering_ECG_CommandOutput}
 \end{align}
 \label{eq:Ctrl_Steering_ECG_Command}
\end{subequations}
%------- Equation -----------------%
\hspace{-0.375cm}
where $ \mu = 1 - \alpha $ and $ 0 \leq \alpha \leq 1 $ is a tuning parameter that corresponds to the time constant of the command virtual dynamics.
If $ \alpha = 0 $, the \ac{ECG} becomes a shift register.

%Figure~\ref{fig:Ctrl_ECG} shows that the \ac{ECG} produces a much more variable command than the standard \ac{LRG}.
%This is mainly due to the method chosen to deal with unfeasible optimization problems (sec.~\ref{sec:Ctrl_Feasibility}).

%\begin{enumerate}
% \item Command Governor activation logic
% 
% \item Command dynamics (when the Command Governor is active)
%
% \item Constraint admissible sets
% 
% \item Command selection method
%\end{enumerate}

%% file: Ctrl_Steering.tex
%=====================================================================================%
\subsection{Steering Control}
\label{sec:Ctrl_Steering}

The output and the constraints of the \ac{RG} are defined as follows:
%------- Equation -----------------%
\begin{subequations}
 \begin{align}
  \mathbf{y} &:= \left[ \begin{array}{cc} LTR & \delta_{SW} \end{array} \right]^{\intercal},\\
  \mathbf{A}_y &:= \left[ \begin{array}{cc}
    1 &  0 \\
   -1 &  0 \\
    0 &  1 \\
    0 & -1 
  \end{array} \right], \: \mathbf{b}_y := \left[ \begin{array}{c}
   LTR_{lim} \\
   LTR_{lim}\\
   \delta_{SW,lim} \\
   \delta_{SW,lim}
  \end{array} \right].
 \end{align}
 \label{eq:Ctrl_Steering_CGDef}
\end{subequations}
%------- Equation -----------------%

The discrete time step for the dynamics matrices is $ \Delta t = 0.01 s $ and the prediction horizon is $ N = 100 $.
For the \ac{ECG}, the optimization gain is $ k_L = 1 $, the virtual state vector size is $ 4 $, and the $ \alpha = 1 - \frac{ \Delta t }{ \tau_{car} } $, so that the virtual dynamics match the vehicle dynamics time constant that, as we have determined empirically, appears to provide best response properties.

%\textbf{ToDo: \emph{Review all parameters.}}

%% file: Ctrl_NonlinearComp.tex
%=====================================================================================%
\subsection{Nonlinear Compensation}
\label{sec:Ctrl_NonlinearComp}

Both the \acl{LRG} and the \acl{ECG} rely on linear model predictions to avoid breaching the defined constraints.
In reality, the controller is acting on a vehicle with nonlinear dynamics.
This results in deviations between the predicted and the real vehicle response.

%=====================================================================================%
\subsubsection{Nonlinear Difference}
\label{sec:Ctrl_NonlinearDiff}

For the same state, there is a difference between the linear model output prediction and the vehicle's real output variables, which we refer to as the nonlinear difference:
%------- Equation -----------------%
\begin{equation}
 \mathbf{d} = f\left(\mathbf{x}, \mathbf{u}\right) - \mathbf{C} \mathbf{x} - \mathbf{D} \mathbf{u} - \mathbf{y}_0.
 \label{eq:Ctrl_NonlinearECG_OutputDiff}
\end{equation}%------- Equation -----------------%
This difference, further exacerbated by the error in the state prediction by the linear model, can either cause the vehicle to breach the constraints when the controller does not predict so, or cause the controller generated command to be too conservative.
This effect can be mitigated if the controller takes into account the current nonlinear difference for the current command computation, assuming that it is persisting over the prediction horizon.
As an example, the nonlinear difference can be compensated in the \ac{LRG} by including it in the $ \mathcal{O}_{ \infty } $ set:
%------- Equation -----------------%
\begin{align}
 \check{\mathcal{O}}_{ \infty } &:= \left\{ \left( \mathbf{v}, \mathbf{x}, \mathbf{d} \right) : \mathbf{y}_k = \right.\notag\\
 &\left. \left( \mathbf{C} \left( \mathbf{I} - \mathbf{A} \right)^{- 1 } \left( \mathbf{I} - \mathbf{A}^k \right) \mathbf{B} + \mathbf{D} \right) \mathbf{v} + \right.\notag\\
 &\left. \mathbf{C} \mathbf{A}^k \mathbf{x} + \mathbf{d} \in \mathbf{Y}, k = 0, ..., N \right\} \bigcap \Gamma'' \subset \mathbb{R}^{4+n+l},
 \label{eq:Ctrl_LRGNC_Oinf}\\
 \Gamma'' &= \left\{ \left( \mathbf{v}, \mathbf{x}, \mathbf{d} \right) | \mathbf{H} \mathbf{v} + \mathbf{d} \in \left( 1 - \epsilon \right) \mathbf{Y} \right\}\notag.
\end{align}
%------- Equation -----------------%

To use the nonlinear difference in the controller, the $ \mathbf{x}_{\mathcal{O}_{ \infty }} $ vector and the $ \mathbf{A}_{\mathcal{O}_{ \infty }} $ matrix are extended to account for $ \mathbf{d} $:
%------- Equation -----------------%
\begin{subequations}
 \begin{align}
  \Breve{\mathbf{x}}_{\mathcal{O}_{ \infty }}^{\intercal} &:= \left[ \begin{array}{ccc} \mathbf{u}^{\intercal}, & \mathbf{x}^{\intercal}, & \mathbf{d}^{\intercal} \end{array} \right]^{\intercal},\\
  \Breve{\mathbf{A}}_{\mathcal{O}_{ \infty }} &:= \left[ \begin{array}{cc}
   \mathbf{A}_{\mathcal{O}_{ \infty }} & \begin{array}{cc}
    \mathbf{A}_y \\
    \mathbf{A}_y \\
    \vdots \\
    \mathbf{A}_y
   \end{array}
  \end{array} \right], \:
  \Breve{\mathbf{b}}_{\mathcal{O}_{ \infty }} := \mathbf{b}_{\mathcal{O}_{ \infty }}.
 \end{align}
 \label{eq:Ctrl_CG_OinfMatrices_Nonlinear}
\end{subequations}
\input{Ctrl_MultiPointLin}

%% file: Ctrl_MultiPointLin.tex
\subsubsection{\ac{MPL}}
\label{sec:Ctrl_MultiPointLin}

%========= Figure ===============%
\begin{figure}[t]
 \centering
 \includegraphics[trim=0.5cm 0.0cm 1.1cm 0.5cm, clip=true, width = 2.2in]{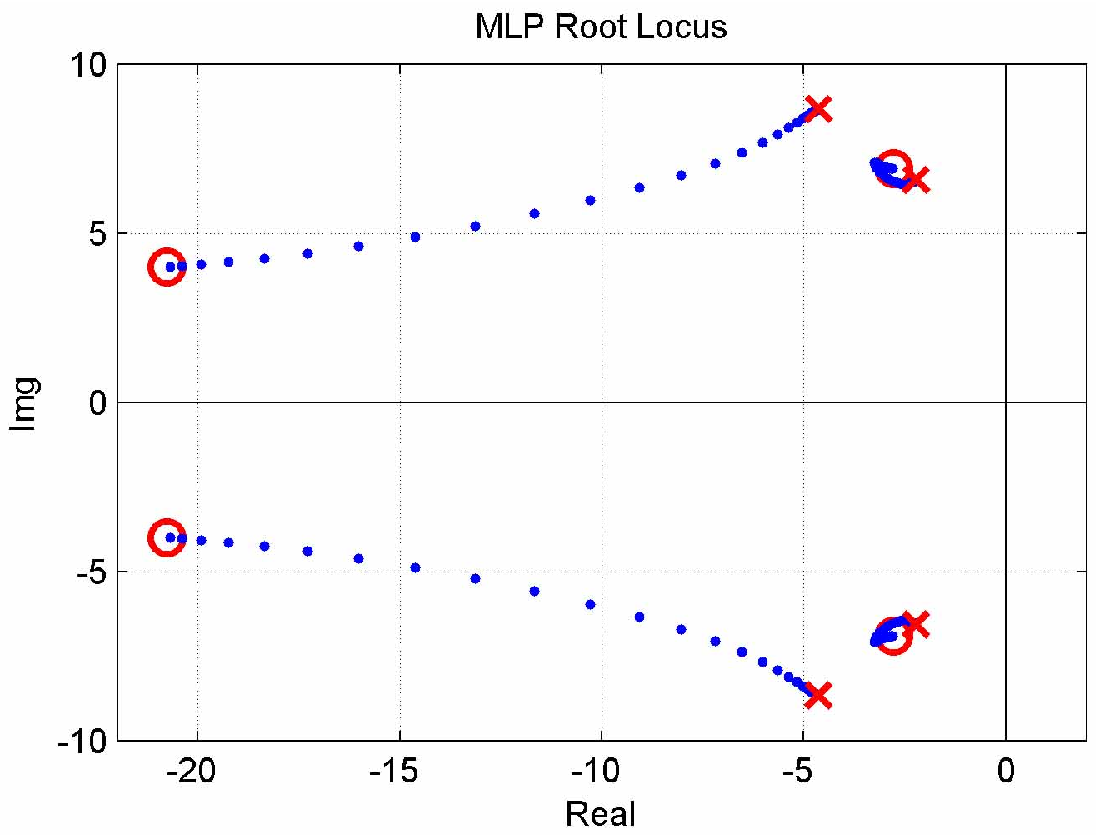}
 \caption{Vehicle dynamics' poles for a range of steering angles from 0 degrees (cross) to 150 degrees (circle).}
 \label{fig:Ctrl_MPL_Poles}
\end{figure}
%========= Figure ===============%

\begin{figure}[t]
 \centering
 \subfloat[Vehicle steering command, \ac{LTR} response, and roll response. The dashed blue line is the reference command. The dot-and-dashed red and the solid black lines are the \ac{LRG} commands with a single- and multi-point linearization, respectively, both with the nonlinear difference compensation.]{
  \includegraphics[trim=0.1cm 0.0cm 0.9cm 0.5cm, clip=true, width = 2.2in]{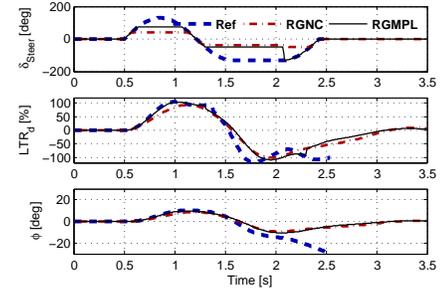}
  \label{fig:Ctrl_MPL_Conservatism}
 }
 \\
 %\hfill
 \subfloat[Linearization point selection in the \ac{LRG} with multi-point linearization.]{
  \includegraphics[width = 2.2in]{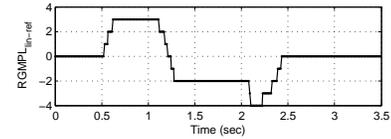}
  \label{fig:Ctrl_MPL_PointSelection}
  }
 \caption{\ac{LRG} with multi-point linearization. Here linearization points used corresponded to steering angles of $ \delta_{SW} = 0, 20, 40, 80 \:\&\: 130 deg $.}
 \label{fig:Ctrl_MPL}
\end{figure}

The nonlinear difference compensation in the previous subsection reduces the nonlinear effects in the vicinity of the system's current state.
However, in this work and in particular for large input commands, that are likely to cause a rollover, this alone is insufficient.
Figure~\ref{fig:Ctrl_MPL_Poles} shows how much the vehicle dynamics' poles can change for a range of steering angles.
If the controller uses a model linearized around the \emph{no actuation} point ($ \delta_{SW} = 0 $), the controller becomes too conservative. %, $ \delta_{DB} = 0 $, $ \Delta K_{ss} = 0 $, $ \Delta D_{ss} = 0 $, $ \overline{ \Delta k }_{ss} = 0 $, and $ \overline{ \Delta d }_{ss} = 0 $
The use of multiple linearization points to define multiple $ \mathcal{O}_{ \infty } $ sets has proved to be an appropriate compensation.
The multi-point linearization results in a less conservative controller, when compared to a controller just with the nonlinear difference compensation (fig.~\ref{fig:Ctrl_MPL_Conservatism}).

The control strategy is the same as described in the previous sections.
The difference is that several linearization points are selected (fig.~\ref{fig:Ctrl_MPL_PointSelection}) and, for each one, an $ \mathcal{O}_{ \infty } $ set \eqref{eq:Ctrl_RG_OinfMatrices} is computed.
The controller then selects the closest linearization point and corresponding $ \mathcal{O}_{ \infty } $ set, based on the current steering angle.
%Additionally, we define threshold points for a variable selected a priori, \emph{e.g.}, $ \delta_{SW} $, the linearization reference variable.
%The controller then selects the appropriate linearization point by considering the thresholds and the present value of the linearization reference variable.
%The controller then interpolates the appropriate linearization points and corresponding $ \mathcal{O}_{ \infty } $ sets based on steering angle to obtain a suitable $ \mathcal{O}_{ \infty } $ set.
Note that in Figure~\ref{fig:Ctrl_MPL_Conservatism} and subsequent figures we report \ac{LTR} in percent.
%========= Figure ===============%
%========= Figure ===============%

%% file: Ctrl_Feasibility.tex
\subsection{Command Computation Feasibility}
\label{sec:Ctrl_Feasibility}

%Due to the mismatch between the vehicle nonlinear dynamics and the controller linear model, t
In practical applications, the \ac{LRG} and \ac{ECG} optimization problems, \eqref{eq:Ctrl_LRG_Cmd} and \eqref{eq:ECG_QP_CostMin}, may become infeasible due to unmodeled uncertainties, in particular, due to the approximation of the nonlinear vehicle dynamics by a linear model in prediction.
Figure \ref{fig:Ctrl_Feas-ExpInterv_Feasibility} shows the computation feasibility classification for an example vehicle simulation.
This section describes the different approaches to deal with infeasibility and the outcomes of their evaluation.

%=====================================================================================%
\subsubsection{Last Successful Command Sequence}
\label{sec:Ctrl_Feasibility_LastSuccessful}

The simplest approach, in the event of infeasibility, is to maintain the last successfully computed command, for the \ac{LRG}:
%------- Equation -----------------%
\begin{equation}
 \mathbf{v}_k = \mathbf{v}_{k - 1},
 \label{eq:Ctrl_Infeas_Maintain_Cmd_LRG}
\end{equation}
%------- Equation -----------------%
or command sequence, for the \ac{ECG}:
%------- Equation -----------------%
\begin{subequations}
 \begin{align}
  \bar{\mathbf{x}}_k &= \bar{\mathbf{A}} \bar{\mathbf{x}}_{k - 1}, \\
  \mathbf{\rho}_k &= \mathbf{\rho}_{ k - 1 }, \\
  \mathbf{v}_k &= \bar{\mathbf{C}} \bar{\mathbf{x}}_k + \mathbf{\rho}_k.
 \end{align}
 \label{eq:Ctrl_Infeas_Maintain_Cmd_ECG}
\end{subequations}
%------- Equation -----------------%
\hspace{-0.25cm}

%=====================================================================================%
\subsubsection{Command Contraction}
\label{sec:Ctrl_Feasibility_Contraction}

%========= Figure ===============%
\begin{figure}[t!]
 \centering
 \subfloat[\scriptsize{Classification of the solution computation feasibility for the \ac{LRG}. Level 1 means the controller is able to compute a solution. Level 0 means that there is no viable solution, because the current output is already breaching the constraints and the current reference is the same as the last used command. The levels $ - 1 $ and $ - 2 $ indicate that for some of the points in the prediction horizon a solution would only exist if the gain $ k_{LRG} $ was set to more than $ 1 $ or less than $ 0 $, respectively. Level $ -3 $ indicates that for different points in the prediction horizon a solution would only exist if the gain $ k_{LRG} $ was set to more than $ 1 $ and less than $ 0 $, simultaneously. The levels $ - 4 $, $ - 5 $ and $ - 6 $ are set when the conditions for level 0 are verified at the same time as those necessary for the levels $ - 1 $, $ - 2 $ and $ - 3 $, respectively. The dashed blue line for the method with command contraction allowed shows when a solution becomes viable due to the contraction.}]{
  \includegraphics[width = 2.2in]{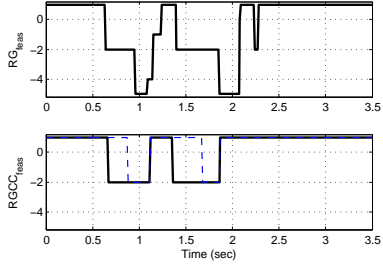}
  \label{fig:Ctrl_Feas-ExpInterv_Feasibility}
 }
 \\
 %\hfill
 \subfloat[Vehicle steering command, \ac{LTR} response, and roll response. The dashed blue line is the reference command. The dot-and-dashed red and the solid black lines are the \ac{LRG} commands with {$ k_{LRG} \in \left[ 0, 1 \right] $} and allowing command contraction, respectively.]{
  \includegraphics[trim=0.1cm 0.0cm 0.9cm 0.5cm, clip=true, width = 2.2in]{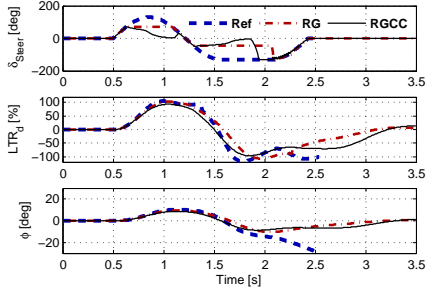}
  \label{fig:Ctrl_Feas-ExpInterv_Dynamics}
 }
 \caption{Feasibility recovery - \ac{LRG} with command contraction allowed.}
 \label{fig:Ctrl_Feas-ExpInterv}
\end{figure}
%========= Figure ===============%
In its standard form, the \ac{LRG} computation limits the \ac{LRG} gain to be $ k_{LRG} \in \left[ 0, 1 \right] $, \emph{i.e.}, the computed command is limited to the interval $ \mathbf{v}_k \in \left[ \mathbf{v}_{k - 1}, \mathbf{u}_k \right] $.
With the \ac{LRG}, there are conditions in which maintaining the last successfully computed command might be problematic.
If the solution infeasibility is caused by the differences between the linear model and vehicle nonlinear dynamics, that usually means that the controller allowed the command to be too large, allowing the vehicle state to go too close to the constraints, and maybe breaching them in the near future (fig.~\ref{fig:Ctrl_Feas-ExpInterv_Dynamics}).
In this case, modifying \eqref{eq:Ctrl_LRG_Cmd} to allow the command to contract even if the reference command is not contracting produces safer solutions. The computations are modified as follows:
%------- Equation -----------------%
\begin{align}
 \mathcal{S} &= \begin{cases}
  \left[ 0, \max \left\{ \mathbf{v}_{k - 1}, \mathbf{u}_k \right\} \right] & \mathbf{v}_{k - 1} > 0 \: \mathsf{and} \: \mathbf{u}_k > 0,\\
  \left[ \min \left\{ \mathbf{v}_{k - 1}, \mathbf{u}_k \right\}, 0 \right] & \mathbf{v}_{k - 1} < 0 \: \mathsf{and} \: \mathbf{u}_k < 0,\\
  \left[ \mathbf{v}_{k - 1}, \mathbf{u}_k \right] & \mathsf{otherwise},
 \end{cases}\\
 \mathbf{v} &= \argmin_{ \mathbf{v} } \left\{ \left| \mathbf{v} - \mathbf{u}_k \right| | \left( \mathbf{v}, \mathbf{x} \right) \in \mathcal{O}_{ \infty } \: \mathsf{and} \: \mathbf{v} \in \mathcal{S} \right\}.
 \label{eq:Ctrl_LRG_ContractCmd}
\end{align}
%------- Equation -----------------%
Figure~\ref{fig:Ctrl_Feas-ExpInterv_Dynamics} shows that with the \emph{last successful command} (RG) the vehicle actually breaches momentarily the imposed constraints, $ LTR < -100\% $ close to $ t = 1.9 s $, while with the contracted command (RGCC) the LTR is kept within the desired bounds and the command converges to reference value sooner ($ @t \approx 1.9 s $).

%%=====================================================================================%
%\subsubsection{\ac{QP} Solution for the \ac{LRG}}
%
%In its standard form, the \ac{LRG} computation limits the computed command to be in the interval $ \mathbf{v}_k \in \left[ \mathbf{v}_{k - 1}, \mathbf{u}_k \right] $.
%If this constraint is dropped, another solution can be searched with \ac{QP}:
%%------- Equation -----------------%
%%\begin{subequations}
% \begin{align}
%  \mathbf{v} &= \argmin_{ \mathbf{v} } \left\{ \frac{1}{2} \mathbf{v}^{\intercal} \mathbf{H} \mathbf{v}' + \mathbf{f} \mathbf{v} | \left( \mathbf{v}, \mathbf{x} \right) \in \mathcal{O}_{ \infty } \right\},
%  \label{eq:Ctrl_Infeas_QP_Cmd_LRG} \\
%  \mathbf{H} &:= \mathbf{Q}, \: \mathbf{f} := - \mathbf{u}_k^{\intercal} \mathbf{Q}, \: \mathbf{Q} := k_L \mathbf{I}, \notag
% \end{align}
%%\end{subequations}
%%------- Equation -----------------%

%=====================================================================================%
\subsubsection{$ \mathcal{O}_{\infty} $ Set Constraints Temporal Removal}

%========= Figure ===============%
\begin{figure}[!t]
 \centering
 \subfloat[Number of rows removed from the \ac{LRG} $ \mathcal{O}_{\infty} $ set.]{
  \includegraphics[width = 2.25in]{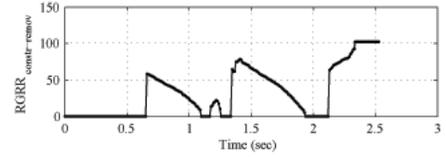}
  \label{fig:Ctrl_Feas-RemoveOinfRows_RemovLin}
 }
 \\
 %\hfill
 \subfloat[Vehicle steering command, \ac{LTR} response, and roll response. The dashed blue line is the reference command. The dot-and-dashed red and the solid black lines are the \ac{LRG} commands blocking and allowing the $ \mathcal{O}_{\infty} $ set rows removal, respectively.]{
 \includegraphics[trim=0.1cm 0.0cm 0.9cm 0.5cm, clip=true, width = 2.2in]{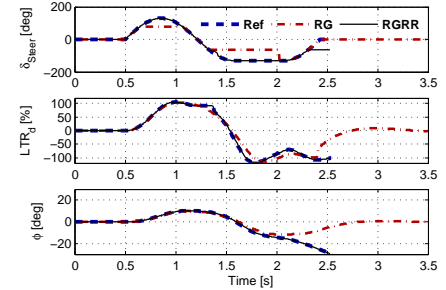}
 \label{fig:Ctrl_Feas-RemoveOinfRows_Dynamics}
 }
 \caption{Feasibility recovery - $ \mathcal{O}_{\infty} $ set rows removal.}
 \label{fig:Ctrl_Feas-RemoveOinfRows}
\end{figure}
%========= Figure ===============%
This method is based on the premise that the controller might not be able to avoid the constraint violation in the near term, but will be able to bring the system into the desired operation envelope in the future.
The method removes as many initial rows from the representation of the $ \mathcal{O}_{\infty} $ set as required to make the command computation feasible (alg.~\ref{alg:Ctrl_Feas-RemoveOinfRows}). 
In Algorithm~\ref{alg:Ctrl_Feas-RemoveOinfRows}, $ \mathbf{A}_{\mathcal{O}_{ \infty }}^{2 i l + 1 : 2 \left( N l + 2 \right) } $ is the matrix composed by all rows between the $ \left( 2 i l + 1 \right)th $ line and the last ($ 2 \left( N l + 2 \right)th $ row) of the $ \mathbf{A}_{\mathcal{O}_{ \infty }} $ matrix.

%------- Algorithm -----------------%
\begin{algorithm}
 $ i = 0 $\\
 \While{$ \mathbf{v} \: computation \: fails $}{
  $ i = i + 1 $\\
  $ \mathbf{A}'_{\mathcal{O}_{ \infty }} = \mathbf{A}_{\mathcal{O}_{ \infty }}^{2 i l + 1 : 2 \left( N l + 2 \right) } $\\
  $ \mathbf{b}'_{\mathcal{O}_{ \infty }} = \mathbf{b}_{\mathcal{O}_{ \infty }}^{2 i l + 1 : 2 \left( N l + 2 \right) } $
 }
 \caption{Constraint removal from the $ \mathcal{O}_{\infty} $ set.}
 \label{alg:Ctrl_Feas-RemoveOinfRows}
\end{algorithm}
%------- Algorithm -----------------%

Figure~\ref{fig:Ctrl_Feas-RemoveOinfRows_RemovLin} indicates the number of rows that had to be removed at each computation step to make the command computation feasible.
As shown by Figure~\ref{fig:Ctrl_Feas-RemoveOinfRows_Dynamics}, this approach (RGRR) is prone to failure, limiting the steering angle only very slightly ($ @t\approx 1.35s $) and allowing the car to rollover.

%=====================================================================================%
\subsubsection{$ \mathcal{O}_{\infty} $ Set Constraints Relaxation}
\label{sec:Ctrl_Feasibility_ConstrRelax}

%========= Figure ===============%
\begin{figure}[b!]
 \centering
 \subfloat[\ac{LRG} constraint relaxation magnitude.]{
  \includegraphics[width = 2.2in]{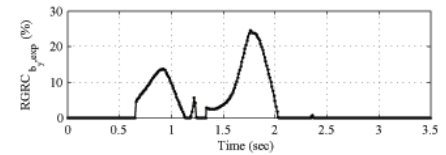}
  \label{fig:Ctrl_Feas-RelaxOinf_ConstrRelax}
 }\\
 \subfloat[Vehicle steering command, \ac{LTR} response, and roll response. The dashed blue line is the reference command. The dot-and-dashed red and the solid black lines are the \ac{LRG} commands without and with constraint relaxation, respectively.]{
   \includegraphics[trim=0.1cm 0.0cm 0.9cm 0.5cm, clip=true, width = 2.2in]{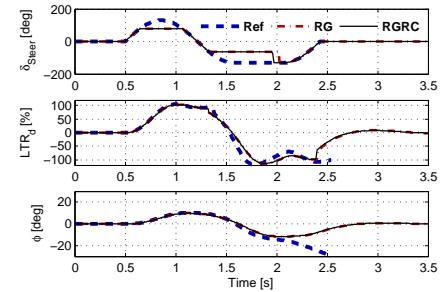}
  \label{fig:Ctrl_Feas-RelaxOinf_Dynamics}
 }
 \caption{Feasibility recovery - $ \mathcal{O}_{\infty} $ constraints relaxation.}
 \label{fig:Ctrl_Feas-RelaxOinf}
\end{figure}
%========= Figure ===============%
The logic behind this approach is that by allowing the controller to find a feasible solution through constraint relaxation, a solution may be found that unfreezes the command computation.
This allows the controller to find a solution that may lead to a behavior closer to the intended than a locked command.
The method expands and contracts the $ \mathcal{O}_{\infty} $ set by modifying the $ \mathbf{b}_{\mathcal{O}_{ \infty }} $ vector:
%------- Equation -----------------%
\begin{equation}
 \mathbf{b}'_{\mathcal{O}_{ \infty }} = \left( 1 + \epsilon \right) \mathbf{b}_{\mathcal{O}_{ \infty }},
 \label{eq:Ctrl_Infeas_bExp_Cmd}
\end{equation}
%------- Equation -----------------%
where $ \epsilon $ is the expansion factor.
This factor is doubled until the command computation is successful and then the bisection method is used to find the minimum $ \epsilon $ for which the command computation is feasible.

Figure~\ref{fig:Ctrl_Feas-RelaxOinf_ConstrRelax} shows the constraint expansion factor that had to be used at each computation step to make the command computation feasible.
Figure~\ref{fig:Ctrl_Feas-RelaxOinf_Dynamics} shows that this method (RGRC) provides only a small improvement over the \emph{last successful command} (RG) by allowing the command to converge to the reference value sooner ($ @t \approx 2.0 s $).
It still allows the LTR to breach the imposed constraints ($ @t \approx 1.9 s $).

We note that various heuristic and sensitivity-based modifications can be proposed where only some of the constraints are relaxed but this entails additional computing effort which is undesirable for this application.
Note also that a soft constraint version of the \ac{RG} has been proposed in \cite{Kalabić2015ReferenceGovernors}; this strategy has not been formally evaluated as it is similar to our $ \mathcal{O}_{ \infty } $ set constraints relaxation approach.

We also note that for the \ac{ECG} a similar constraint relaxation method could be implemented, where a relaxation variable is included as part of the \ac{QP} and the constraints that are most violated are relaxed first.

%=====================================================================================%
\subsubsection{Selected Feasibility Recovery Method}

From the different methods tested with the \ac{LRG}, the command contraction method was the best performing  (sec.~\ref{sec:Results_CtrlComp_NoError} and fig.~\ref{fig:Results_FeasRecovery}).
In the simulations section, except for the results comparing directly the infeasibility recovery methods, the \ac{LRG} is tested with the command contraction method.

%% file: Ctrl_NRG.tex
%=====================================================================================%
\subsection{\acf{NRG}}
\label{sec:Ctrl_NonlinearRG}

%========= Figure ===============%
\begin{figure}[b!]
 \centering
 \subfloat[Steering command.]{
  \includegraphics[width = 2.2in]{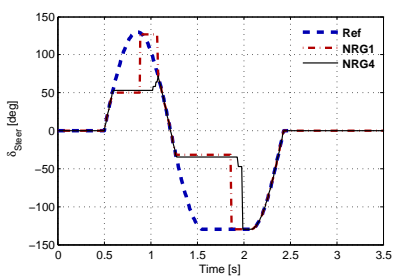}
  \label{fig:Ctrl_NRG_NPred_Command}
 }
 \\
 %\hfill
 \subfloat[Vehicle lateral response.]{
  \includegraphics[width = 2.2in]{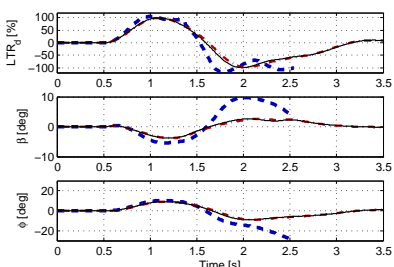}
  \label{fig:Ctrl_NRG_NPred_Dynamics}
 }
 \caption{\acl{NRG} command with 1 and 4 nonlinear iterations per step.}
 \label{fig:Ctrl_NRG_NPred}
\end{figure}
%========= Figure ===============%
% Pseudo-code
The \ac{NRG} relies on a nonlinear model in prediction to check if a command is safe or, if otherwise, compute a safe command.
Instead of the $ \mathcal{O}_{ \infty } $ set used by the \acp{LRG}, the \ac{NRG} uses a nonlinear model (sec.~\ref{sec:Model_NonlinearCar_BodyEOM}) to predict the vehicle response to a constant command for the specified time horizon, usually comparable to the settling time.
If the predicted vehicle response stays within the imposed constraints, the command is deemed safe and is passed on.
%The \ac{NRG} only passes to the actuators a command deemed safe.
%If a reference command results in a predicted constraint breach, the \ac{NRG} uses a bisection iterative method to find a safe command in the range between the last passed command, a known safe command, and the reference command.
Otherwise, the \ac{NRG} uses bisections to find a safe command in the range between the last passed (safe) command and the reference command.
Each iteration, including the first with the reference command, involves a nonlinear prediction of a modified by bisections command, checks if it respects the constraints, and classifies the command as safe or unsafe.
These bisection iterations numerically minimize the reference-command difference, \emph{i.e.}, the difference between the reference command and the used safe command.
The number of iterations is a configuration parameter, and governs the balance between the computation time and the solution accuracy.
Parametric uncertainties can be taken into account following the approach in \cite{Sun2005ReferenceGovernors}, but these developments are left to future work.

Figure~\ref{fig:Ctrl_NRG_NPred} shows that the \ac{NRG} with a single iteration (NRG1), \emph{i.e.}, a simple verification of the trajectory constraints with the reference command, produces very similar results when compared with the \ac{NRG} with three extra bisections (NRG4).
The NRG4 initially allows a slightly less constrained command and then has slightly smoother convergence with the reference command ($ @t = 0.8 s \:\&\: 1.7 s $).
This behavior is obtained at the expense of the computational load, taking about 4 times longer to compute a safe command when the reference is unsafe.
For the illustrated example (fig.~\ref{fig:Ctrl_NRG_NPred}), the NRG1 and the NRG4 take an average of $ 0.16 s $ and $ 0.31 s $, respectively, to compute a safe command.

%% file: Results_SimSetup.tex
\subsection{Simulation Setup}
\label{sec:Results_SimSetup}

%========= Figure ===============%
\begin{figure*}[t]
 \centering
 \includegraphics[width = 6in]{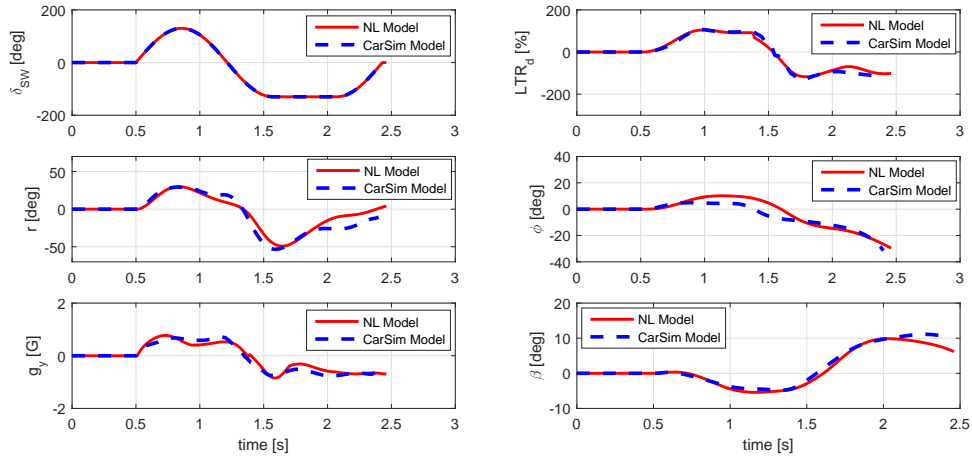}
 \caption{Vehicle dynamics in a Sine with Dwell maneuver. Comparison between the trajectories simulated by a CarSim model and the nonlinear model (sec.~\ref{sec:Model_NonlinearCar}).}%, as well as the trajectories generated with the intervention of the proposed command governors
 \label{fig:Results_CarSimComparison}
\end{figure*}
%========= Figure ===============%
%========= Figure ===============%
\begin{figure*}[!t]
 \centering
 \includegraphics[width = 6in]{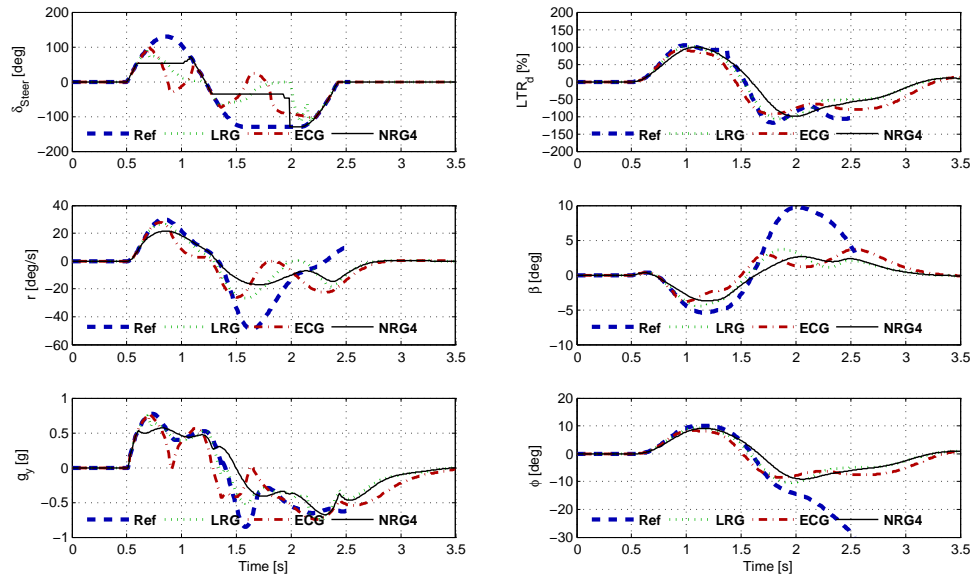}
 \caption{Vehicle dynamics in a Sine with Dwell maneuver. Comparison between the trajectories generated with the intervention of the proposed reference governors.}
 \label{fig:Results_TrajComparison}
\end{figure*}
%========= Figure ===============%

%=====================================================================================%
\subsubsection{Simulation Model}

The nonlinear simulation model was setup to have a behavior similar to a North American SUV.
The model parameters are listed in Table~\ref{tab:Results_Vehicle_SimParam}.

%---------- Table ---------------%
\begin{table}[h!]
 \footnotesize
 \caption{Vehicle simulation parameters.}
 \label{tab:Results_Vehicle_SimParam}
 \begin{centering}
  \begin{tabular}{|l||c|}
   \hline Parameter & Value \tabularnewline
   \hline
   \hline $ l_f $ & $ 1.160 m $ \tabularnewline
   \hline $ l_r $ & $ 1.750 m $ \tabularnewline
   \hline $ T $ & $ 1.260 m $ \tabularnewline
   \hline $ h_{SM} $ & $ 0.780 m $ \tabularnewline
   \hline $ h_{UC} $ & $ 0.000 m $ \tabularnewline
   \hline $ m $ & $ 2000 kg $ \tabularnewline
   \hline $ m_{SM} $ & $ 1700 kg $ \tabularnewline
   \hline $ m_{UC} $ & $ 300 kg $ \tabularnewline
   \hline $ I_{xx,SM} $ & $ 1280 kg/m^{2} $ \tabularnewline
   \hline $ I_{xx,UC} $ & $ 202 kg/m^{2} $ \tabularnewline
   \hline $ I_{yy,SM} $ & $ 2800 kg/m^{2} $ \tabularnewline
   \hline $ I_{zz} $ & $ 2800 kg/m^{2} $ \tabularnewline
   \hline $ I_{xz,SM} $ & $ 0 kg/m^{2} $ \tabularnewline
   \hline $ k_{\delta_{SM}} $ & $ 17.5 $ \tabularnewline
   \hline $ K_{S} $ & $ 73991 N.m $ \tabularnewline
   \hline $ D_{S} $ & $ 5993 N.m.s/rad $ \tabularnewline
   \hline $ \overline{ \Delta k }_{ss} $ & $ 0.0 \% $ \tabularnewline
   \hline $ \overline{ \Delta d }_{ss} $ & $ 0.0 \% $ \tabularnewline
%   \hline $ K_{S,xx} $ & $ 73991 N.m $ \tabularnewline
%   \hline $ D_{S,xx} $ & $ 5993 N.m.s/rad $ \tabularnewline
%   \hline $ K_{S,yy} $ & $ 170880 N.m $ \tabularnewline
%   \hline $ D_{S,yy} $ & $ 13842 N.m.s/rad $ \tabularnewline
   \hline
  \end{tabular}
  \par
 \end{centering}
\end{table}
%---------- Table ---------------%

We used a CarSim\textsuperscript{\textregistered} simulation model to check the realism of the nonlinear model presented in section~\ref{sec:Model_NonlinearCar}, implemented in MATLAB\textsuperscript{\textregistered}.
Figure~\ref{fig:Results_CarSimComparison} illustrates the simulation results from both models in terms of their lateral dynamics.
The lateral dynamics match very well.
All variables except for the roll angle match both in trend and amplitude.
The roll angle matches in trend, but shows a larger amplitude in our MATLAB\textsuperscript{\textregistered} model.
More importantly, both simulations match very well in terms of the main rollover metric, the \ac{LTR}.
The simulations diverge more in the last moments, when the roll angle is increasing and the car is in the process of rolling over.

%=====================================================================================%
\subsubsection{Test Maneuvers}
\label{sec:Results_TestMan}

%%========= Figure ===============%
%\begin{figure}[!t]
% \centering
% \subfloat[Steering command.]{
%  \includegraphics[height = 2.2in]{TestMan_Paper_Cmd_v1_02}
%  \label{fig:Results_TestMan_Command}
% }
% \\%\hfill
% \subfloat[Vehicle response.]{
%  \includegraphics[height = 2.2in]{TestMan_Paper_Ctrl_v1_02}
%  \label{fig:Results_TestMan_Response}
% }
% \caption{Vehicle stability test maneuvers: \emph{Sine with Dwell, J-Turn, and FishHook}.}
% \label{fig:Results_TestMan}
%\end{figure}
%%========= Figure ===============%
\ac{NHTSA} defines several test maneuvers: Sine with Dwell, J-Turn, and FishHook \cite{NHTSA2003Rollover}.
In this work, we chose to test the controllers and demonstrate rollover avoidance for Sine with Dwell maneuvers (fig.~\ref{fig:Results_CarSimComparison}).
%========= Figure ===============%
\begin{figure}[!t]
 \centering
 \subfloat[Sprung mass and undercarriage maximum roll angles.]{
  \includegraphics[width = 2.2in]{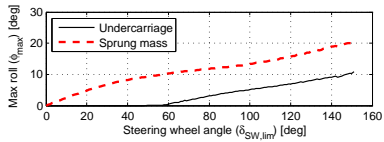}
  \label{fig:Results_WheelLift_Roll}
 }
 \\%\hfill
 \subfloat[Maximum wheel lift.]{
 \includegraphics[width = 2.2in]{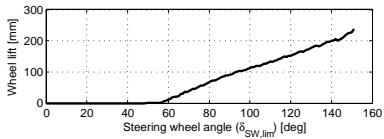}
  \label{fig:Results_WheelLift_Lift}
 }
 \caption{Variation of the roll propensity for Sine with Dwell maneuvers. The Sine with Dwell maneuvers vary in terms of steering amplitude, defined by the limit steering angle ($ \delta_{SW,lim} $).}
 \label{fig:Results_WheelLift}
\end{figure}
%========= Figure ===============%
%The maximum wheel lift attained during a maneuver is obtained as: $ h_{WL,max} = T \sin \left| \phi_{uc,max} \right| $.
Figure~\ref{fig:Results_WheelLift_Lift} illustrates the variation of the vehicle roll (spring mass and undercarriage) and of the maximum wheel lift ($ h_{WL,max} = T \sin \left| \phi_{uc,max} \right| $) with respect to the maximum value of the Sine with Dwell reference steering angle, showing $ \approx 20 deg $ of sprung mass roll and $ \approx 240 mm $ of wheel lift for $ \delta_{SW,lim} = 150 deg$.

%% file: Results_RG_Steering.tex
%=====================================================================================%
\subsection{Trajectories with the \acfp{RG}}
\label{sec:Results_RG_Steering}

%%========= Figure ===============%
%\begin{figure*}[!t]
% \centering
% \includegraphics[width = 6in]{CtrlComp-LatDynDet}
% \caption{Vehicle dynamics in a Sine with Dwell maneuver. Comparison between the trajectories generated with the intervention of the proposed command governors.}
% \label{fig:Results_TrajComparison}
%\end{figure*}
%%========= Figure ===============%
%========= Figure ===============%
\begin{figure}[!b]
 \centering
 \includegraphics[width = 2.2in]{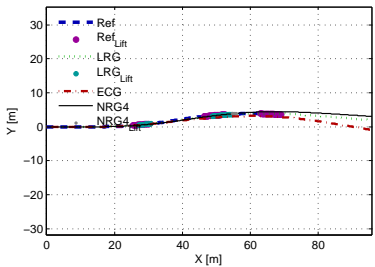}
 \caption{Vehicle trajectory in a \emph{Sine with Dwell} maneuver.}
 \label{fig:Results_RG_Steering_SineDwell_Traj}
\end{figure}
%========= Figure ===============%
%%========= Figure ===============%
%\begin{figure}[!b]
% \centering
% \subfloat[Steering command.]{
%  \includegraphics[width = 2.2in]{CtrlComp_Cmd}
%  \label{fig:Results_RG_Steering_SineDwell_Command}
% }
% \\
% %\hfill
% \subfloat[Vehicle trajectory.]{
%  \includegraphics[width = 2.2in]{CtrlComp_Pos}
%  \label{fig:Results_RG_Steering_SineDwell_Traj}
% }
% \caption{Stability test with Reference Governors: \emph{Sine with Dwell}.}
% \label{fig:Results_RG_Steering_SineDwell}
%\end{figure}
%%========= Figure ===============%
%========= Figure ===============%
\begin{figure}[!b]
 \centering
 \includegraphics[width = 2.2in]{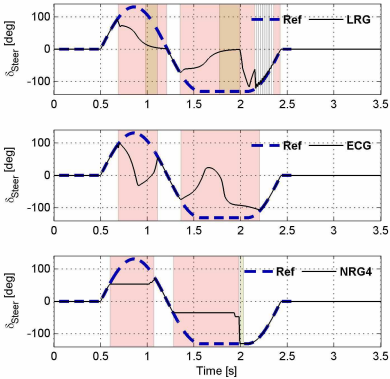}
 \caption{Reference Governors activation, \emph{i.e.}, steering command modification.}
 \label{fig:Results_RG_Steering_SineDwell_CtrlActivation}
\end{figure}
%========= Figure ===============%

Figures~\ref{fig:Results_TrajComparison} and \ref{fig:Results_RG_Steering_SineDwell_Traj} illustrate the effect of the various \acp{RG} on the vehicle trajectory.
The simulations have been performed on our nonlinear vehicle dynamics model in MATLAB\textsuperscript{\textregistered}.
The \acf{LRG} used in this comparison uses the \emph{nonlinear difference} compensation (sec.~\ref{sec:Ctrl_NonlinearDiff}), the \emph{\acf{MPL}} (sec.~\ref{sec:Ctrl_MultiPointLin}), and allows command contraction (sec.~\ref{sec:Ctrl_Feasibility_Contraction}). % when an infeasible command computation is found.
The \acf{ECG} uses the \emph{\ac{MPL}} (sec.~\ref{sec:Ctrl_MultiPointLin}) and maintains the last successfully computed command sequence (sec.~\ref{sec:Ctrl_Feasibility_LastSuccessful}) when an infeasible command computation is found.
The \acf{NRG} performs 4 iterations to find a suitable command, when the reference command is deemed unsafe.

It is clear from Figure~\ref{fig:Results_TrajComparison} that the \acp{RG} steering adjustments are different in shape, but not so much in their effect on the \ac{LTR}, side-slip ($ \beta $), roll ($ \phi $), and lateral acceleration ($ g_y $).
The amplitude of the turn rate ($ r $) is more limited by the \ac{NRG} than the other \acp{RG}.
The trajectories with the \acp{RG} do not diverge much from the trajectory with the reference steering command, up to the moment when the vehicle starts to rollover with the reference command (fig.~ \ref{fig:Results_RG_Steering_SineDwell_Traj}).
The $ Ref_{Lift} $, $ LRG_{Lift} $, and $ NRG4_{Lift} $ shades over the x-y trajectory illustrate where the \ac{LTR} breaches the imposed constraints.

Figure~\ref{fig:Results_RG_Steering_SineDwell_CtrlActivation} highlights with background shades when the \acp{RG} are active, adjusting the steering command.
The darker shades indicate that the controller was unable to avoid breaching the \ac{LTR} constraints.
The \ac{LTR} plots in Figures~\ref{fig:Results_TrajComparison} and \ref{fig:Results_RG_Steering_SineDwell_CtrlActivation} show that in this simulation instance the \ac{LRG} allowed the \ac{LTR} to slightly exceed the constraints ($ @ t \approx 1 s \:\&\: 1.8 s $), but was able to maintain the roll angle well under control.

%% file: Results_CtrlComp_NoError.tex
%=====================================================================================%
\subsection{\aclp{RG} Performance Comparison}
\label{sec:Results_CtrlComp_NoError}

%========= Figure ===============%
\begin{figure}[!b]
 \centering
 \subfloat[Steering command.]{
  \includegraphics[width = 2.2in]{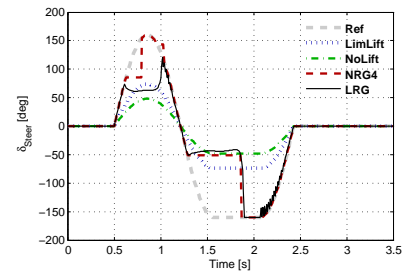}
  \label{fig:Results_RG_TestComparison_Command}
 }
 \\%\hfill
 \subfloat[Vehicle trajectory on the X-Y plane.]{
  \includegraphics[width = 2.2in]{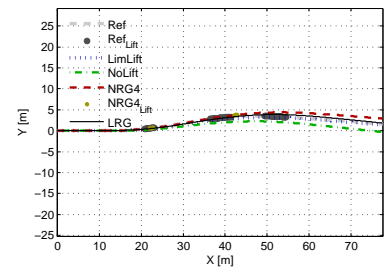}
  \label{fig:Results_RG_TestComparison_Traj}
 }
 \caption{Reference safe trajectories used for the reference governors performance evaluation.
  The \emph{Ref} trajectory is the unmodified sine-with-dwell maneuver with the maximum steering angle defined for a specific test, in this case \emph{160 degrees}.
  The \emph{LimLift} trajectory is the sine-with-dwell maneuver that produces the maximum allowable wheel lift (\emph{5 cm, 2"}).
  The \emph{NoLift} trajectory is the sine-with-dwell maneuver with the maximum steering angle that produces no wheel lift.
  The \emph{NRG4} trajectory illustrates the \emph{safe} maneuver, resulting from the Nonlinear Reference Governor intervention, that is considered to interfere the least with the original trajectory (\emph{Ref}), while avoiding almost completely wheel lift conditions.
  The \emph{LRG} trajectory illustrates a maneuver resulting from the application of the Linear Reference Governor.}
 \label{fig:Results_RG_TestComparison}
\end{figure}
%========= Figure ===============%

The results presented next characterize the performance of the controllers in terms of constraint enforcement \emph{effectiveness}, the adherence to the driver reference (\emph{conservatism}), the adherence to the desired turning rate (\emph{turning response}), and the controllers' computation time.
The results were obtained from simulation runs with a range of Sine-with-Dwell maneuvers' amplitudes between 10 and 160 deg. %: $ \delta_{Steer} \in \left[ 10, 160 \right] deg $.

Figure \ref{fig:Results_RG_TestComparison} illustrates the trajectories that serve as reference in the \acp{RG}' performance evaluation and an example of a trajectory with a controller intervention (\ac{LRG}).
The reference safe trajectories used in \eqref{eq:Model_Conservatism} and \eqref{eq:Model_TurnResp} are: the reference trajectory, produced by the original command; the limit lift trajectory (\emph{LimLift}), with the maximum allowable wheel lift (\emph{5 cm, 2"}); the no lift trajectory (\emph{NoLift}), that produces no wheel lift; and the quasi-optimal safe trajectory (\emph{NRG4}), produced by an \ac{NRG} with 4 iterations. %, considered to interfere the least with the original trajectory (\emph{Ref}), while avoiding almost completely wheel lift conditions
Each reference safe trajectory has its own merits in the evaluation of the reference governors' conservatism.
The trajectory produced by the reference governors will be the least conservative when compared with the \emph{NoLift} trajectory.
This shows how much the controller reduces the conservatism when compared to a simplistic safe trajectory.
The comparison with the \emph{NRG4} trajectory produces a middle range conservatism evaluation, allowing us to compare the reference governors' command with an almost optimal constraint enforcement strategy.
The reference governors' trajectory will be the most conservative when compared with the \emph{LimLift} trajectory.
This shows how much leeway exists between the reference governors' commands and the commands that produce the limit lift condition.

%========= Figure ===============%
\begin{figure}[!b]
 \centering
 \includegraphics[width = 2.2in]{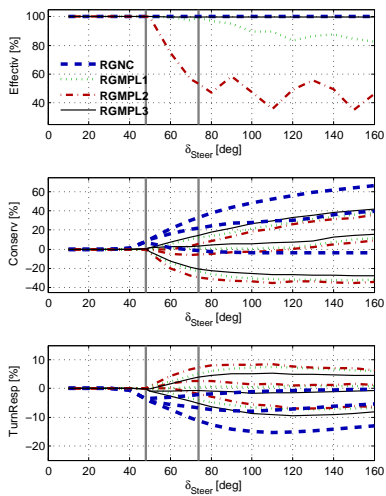}
 \caption{Variation in performance of the \ac{LRG} with a single linearization point (RGNC - \ac{LRG} with just the nonlinear difference compensation) and different selections of MPLs (RGMPL\#).}
 \label{fig:Results_MPLSelection}
\end{figure}
%========= Figure ===============%

The two bottom plots in Figure~\ref{fig:Results_MPLSelection} illustrate the comparison between different \ac{LRG} options and the \emph{NoLift}, \emph{NRG4}, and \emph{LimLift} reference safe trajectories.
On all the figures that illustrate the conservatism and turning response for the same controller configuration, there are three comparison branches, where the \emph{NoLift} and the \emph{LimLift} trajectories offer the most advantageous and most disadvantageous comparison trajectories, respectively.
That means that the \emph{NoLift} branch is the bottom branch in the conservatism plots and the top one in the turning response plots (fig.~\ref{fig:Results_MPLSelection}).
The middle branch, the comparison with a trajectory considered to be \emph{NRG4}, is the most interesting, as it shows a comparison with one of the least conservative trajectories that produce almost no wheel lift. %, and does not constrain the steering command to have the same shape as the original command.

%\textbf{ToDo: \emph{Figure showing the variation in performance of the standard \ac{LRG} for different choices of $ LTR_{max} $ with respect to the maximum value of the maneuver reference command.}}
%%========= Figure ===============%
%\begin{figure}[!t]
% \includegraphics[width = 2.2in]{LTRMax_RG-CtrlPerfTime}
% \caption{Variation in performance of the standard \ac{LRG} for different choices of $ LTR_{max} $.}
% \label{fig:Results_LTRMax}
%\end{figure}
%%========= Figure ===============%

Figure~\ref{fig:Results_MPLSelection} shows how the \ac{LRG} performance improves with Multi-Point Linearization (MPL) and how it changes with the selection of linearization points:
\begin{itemize}
 \item \emph{RGMPL1}: 0, 20, 40, and 100 $ deg $;
 \item \emph{RGMPL2}: 0, 80, 110, and 150 $ deg $;
 \item \emph{RGMPL3}: 0, 20, 40, 60, 80, 100, 120, 130, 140, and 150 $ deg $.
\end{itemize}
Note that fewer linearization points might lead to less conservatism and a better turning response, however may also result in a worse effectiveness (fig.~\ref{fig:Results_MPLSelection}).
With a dense selection of linearization points (RGMPL3 case), the \ac{LRG} is $ 100\% $ effective for all amplitudes of the reference command.

%\textbf{ToDo: \emph{Figure showing the variation in performance of the best MPL \ac{LRG} allowing contraction with different contraction time constants with respect to the maximum value of the maneuver reference command.}}

%========= Figure ===============%
\begin{figure}[!b]
 \centering
 \includegraphics[width = 2.2in]{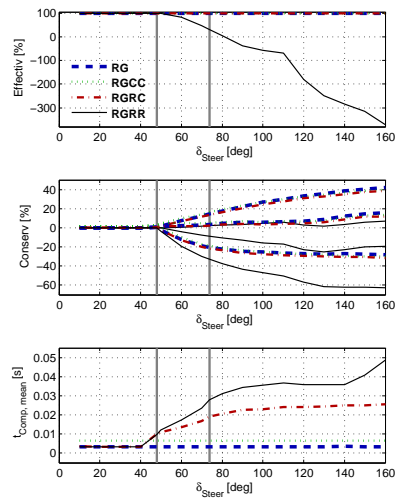}
 \caption{Variation in performance of the MPL \ac{LRG} (RGMPL3's linearization points) with the different feasibility recovery methods: last successful command (\emph{RG}), command contraction (\emph{RGCC}), constraints' relaxation (\emph{RGRC}), and constraint temporal removal (\emph{RGRR}).}
 \label{fig:Results_FeasRecovery}
\end{figure}
%========= Figure ===============%
The feasibility recovery methods are compared with the linearization points of the RGMPL3 case (fig.~\ref{fig:Results_FeasRecovery}).
From the various feasibility recovery methods presented in Section~\ref{sec:Ctrl_Feasibility}, the \emph{constraint temporal removal} (\emph{RGRR}), where rows from the $ \mathcal{O}_{ \infty } $ are removed, is the worst performing.
The other three methods, \emph{last successful command} (\emph{RG}), \emph{command contraction} (\emph{RGCC}), and \emph{constraints relaxation} (\emph{RGRC}), are similar in terms of the conservatism metric.
The \emph{last successful command} (\emph{RG}) method requires the lowest computational overhead.
Nevertheless, the \emph{command contraction} (\emph{RGCC}) method is preferred, because it provides better effectiveness than the other three methods when the state estimation includes some noise (sec.~\ref{sec:Results_CtrlComp_EstimError}).

%========= Figure ===============%
\begin{figure}[!t]
 \centering
 \includegraphics[width = 2.2in]{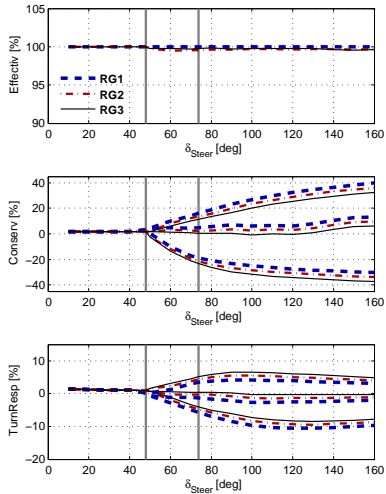}
 \caption{Variation in performance of the MPL \ac{LRG} allowing contraction for different choices of $ LTR_{max} $. $ LTR_{max} = 99, \: 102, \:\&\: 105 \% $ for \emph{RG1}, \emph{RG2}, and \emph{RG3}, respectively.}
 \label{fig:Results_LTRMax}
\end{figure}
%========= Figure ===============%
Most of the \acp{RG}' tests shown were run with an \ac{LTR} constraint of $ 0.99 $ or $ 99\% $.
The \ac{LTR} constraint upper bound can be relaxed, as shown in Figure~\ref{fig:Results_LTRMax}, to reduce the controllers' conservatism.
For $ LTR_{max} = 99, \: 102, \:\&\: 105 \% $, the effectiveness is only slightly degraded and the conservatism is reduced about $ 10\% $ from $ LTR_{max} = 99 \% $ to $ LTR_{max} = 105 \% $.
Note that the relaxation of the \ac{LTR} constraint beyond $ 100\% $ is an ad hoc tuning method, without any guarantees on the effectiveness performance.

%\textbf{ToDo: \emph{Figure showing the variation in performance of the best MPL \ac{ECG} for different eigenvalue selections with respect to the maximum value of the maneuver reference command.}}
%%========= Figure ===============%
%\begin{figure}[!t]
% \centering
% \includegraphics[width = 2.2in]{ECGTConst-CtrlPerfTime}
% \caption{Variation in performance of the best MPL \ac{ECG} for different selections of virtual dynamics time constant.}
% \label{fig:Results_ECGTConst}
%\end{figure}
%%========= Figure ===============%

%========= Figure ===============%
\begin{figure}[!t]
 \centering
 \includegraphics[width = 2.2in]{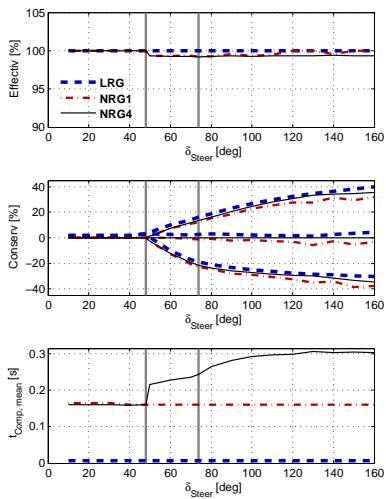}
 \caption{Variation in performance of the \ac{LRG}, NRG1, and NRG4.}
 \label{fig:Results_NRGNPred}
\end{figure}
%========= Figure ===============%
As expected, the \ac{NRG} for a single iteration (\emph{NRG1}), \emph{i.e.}, with a check of the actual reference command, runs faster than the \ac{NRG} setup for four iterations (\emph{NRG4}) (bottom plot of fig.~\ref{fig:Results_NRGNPred}).
The unexpected result is that the \emph{NRG1} setup is less conservative and has higher effectiveness (top plots of fig.~\ref{fig:Results_NRGNPred}).
This happens with Sine with Dwell maneuver, because the \emph{NRG1} is slightly more conservative during the increment of the reference command, i.e., the moment at which the \ac{NRG} command diverges from the reference command, but that produces an earlier convergence with the reference command (fig.~\ref{fig:Ctrl_NRG_NPred}), while the \emph{NRG4} takes longer to converge, with a much larger difference between the reference command and the \ac{NRG} command.

%========= Figure ===============%
\begin{figure}[t]
 \centering
 \includegraphics[width = 2.2in]{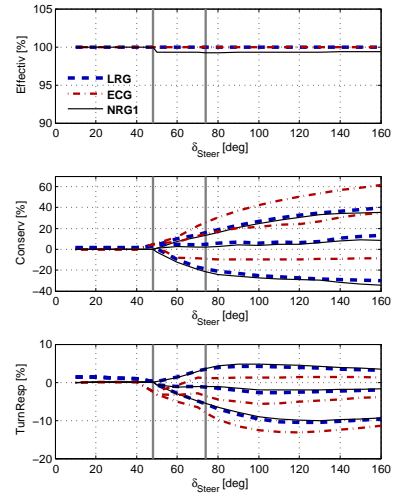}
 \caption{\acp{RG}' performance without estimation errors.}
 \label{fig:Results_MonteCarlo_NoError_CtrlPerf}
\end{figure}
%========= Figure ===============%
%%========= Figure ===============%
%\begin{figure}[t]
% \centering
% \includegraphics[width = 2.2in]{RGComp-CtrlPerf_CompTime}
% \caption{\acp{RG}' computation time without estimation errors.}
% \label{fig:Results_MonteCarlo_NoError_CompTime}
%\end{figure}
%%========= Figure ===============%
%Figures \ref{fig:Results_MonteCarlo_NoError_CtrlPerf} and \ref{fig:Results_MonteCarlo_NoError_CompTime} illustrate the performance of the \ac{LRG},  \ac{ECG}, and \ac{NRG}, for the selected configurations.
Figure \ref{fig:Results_MonteCarlo_NoError_CtrlPerf} illustrates the performance of the \ac{LRG},  \ac{ECG}, and \ac{NRG}, for the selected configurations.
The effectiveness is very similar and over $ 99 \% $ for all the \acp{RG} (top plot of fig.~\ref{fig:Results_MonteCarlo_NoError_CtrlPerf}).
That means that the \acp{RG} keep the wheels from lifting more than $ 0.5 mm $ ($ 0.02 " $) from the ground, even when the reference steering would lead to rollover.
The \ac{LRG} and the \ac{NRG} are less conservative and provide better turn response than the \ac{ECG} (bottom plots of fig.~\ref{fig:Results_MonteCarlo_NoError_CtrlPerf}).
For the most demanding conditions (larger reference command amplitudes), both the \ac{LRG} and the \ac{ECG} have a mean computation time of about $ 0.005 s $.
In the same conditions, the maximum time for the command computation was lower than $ 0.01 s $ for the \ac{LRG} and about $ 0.02 s $ for the \ac{ECG}.
The \ac{NRG} has a larger computation time, which is about $ 0.16 s $ per command update step, when the simulation and control update step is $ 0.01 s $.
This means that the \ac{NRG} setup tested is not able to compute the control solution in real-time in MATLAB\textsuperscript{\textregistered} in the computer used for these tests (64-bit; CPU: Intel\textsuperscript{\textregistered} Core\texttrademark i7-4600U @ 2.70 GHz; RAM: 8 GB).
In C++, the \ac{NRG} would be about 10 times faster.
Also, it may be possible to use slower update rates or shorter prediction horizons with NRGs to reduce the computation times, provided this does not cause performance degradation or increase in constraint violation.
We note that explicit reference governor \cite{Garone2016ReferenceGovernors} cannot be used for this application as we are lacking a Lyapunov function.

%Also the \ac{NRG} could run in real-time for slower control update rates.
%Research is needed to determine the slowest acceptable control update rate, \emph{i.e.}, to characterize how slower update rates degrade the control performance and the driver handling quality perception.

%Figure \ref{fig:Results_MonteCarlo_NoError_CompTime} illustrates the variation in the controllers computation time, showing the mean, the standard deviation, the minimum, and the maximum computation time for each update step.
%The simulation update step is $ 0.01 s $, meaning that the \ac{NRG} is not able to compute the control solution in real-time in the computer used for these tests (64-bit; CPU: Intel\textsuperscript{\textregistered} Core\texttrademark i7-4600U @ 2.70 GHz; RAM: 8 GB).

%% file: Results_CtrlComp_EstimError.tex
%=====================================================================================%
\subsection{Performance in the Presence of Estimation Error}
\label{sec:Results_CtrlComp_EstimError}

%Monte Carlo sampling results on the different controllers and different input errors:
%\begin{enumerate}
% \item Effectiveness;
% 
% \item Conservatism.
%\end{enumerate}

%%=====================================================================================%
%\subsubsection{Control with Perfect Measurements}
%\label{sec:Results_MonteCarlo_PerfectMeas}
%
%\textbf{ToDo: \emph{Figure showing the variation of the effectiveness with respect to the maximum value of the maneuver reference command.}}
%
%\textbf{ToDo: \emph{Figure showing the variation of the conservatism with respect to the maximum value of the maneuver reference command.}}
%
%\textbf{ToDo: \emph{Figure showing the computations power required for each one of the methods with respect to the maximum value of the maneuver reference command.}}
%
%As expected the 
%
%%=====================================================================================%
%\subsubsection{Control with Estimation Errors}
%\label{sec:Results_MonteCarlo_EstimationErrors}

%\textbf{ToDo: \emph{Figure showing the variation of the mean, minimum, and maximum for the effectiveness with the maximum of the maneuver reference command.}}

%\textbf{ToDo: \emph{Figure showing the variation of the mean, minimum, and maximum for the conservatism with the maximum of the maneuver reference command.}}

%\textbf{ToDo: \emph{Figure showing the computations power required for each one of the methods with respect to the maximum value of the maneuver reference command.}}

In this section, we illustrate the variation of the control performance for 3 controllers: a \ac{LRG} with command contraction, an \ac{ECG}, and a \ac{NRG} with a single iteration (\emph{NRG1}).
The controllers are evaluated through Monte Carlo sampling with a range of estimation error conditions in all states used by the controllers: side-slip, turn rate, roll angle, and roll rate.

%========= Figure ===============%
\begin{figure}[t]
 \centering
 \includegraphics[width = 2.2in]{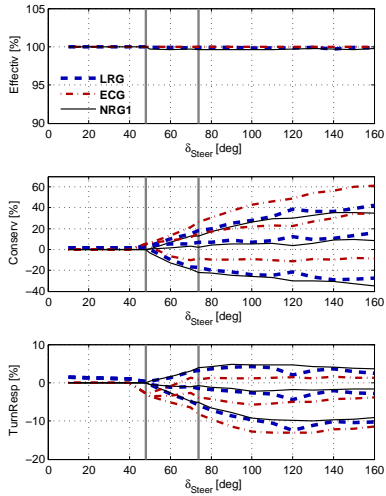}
 \caption{\acp{RG}' performance with estimation errors of $ \sigma = 10\% $ about the true roll angle.}
 \label{fig:Results_MonteCarlo_RollError010_CtrlPerf}
\end{figure}
%========= Figure ===============%
%========= Figure ===============%
\begin{figure}[t]
 \centering
 \includegraphics[width = 2.2in]{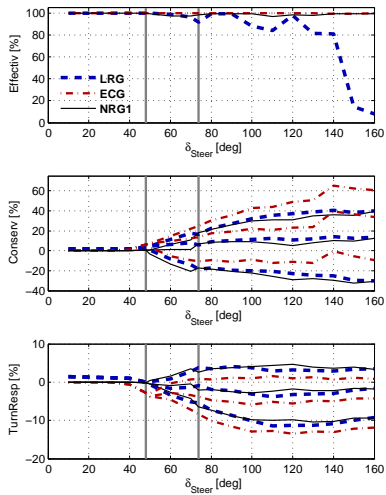}
 \caption{\acp{RG}' performance with estimation errors of $ \sigma = 20\% $ about the true roll angle.}
 \label{fig:Results_MonteCarlo_RollError_CtrlPerf}
\end{figure}
\begin{figure}[t]
 \centering
 \includegraphics[width = 2.2in]{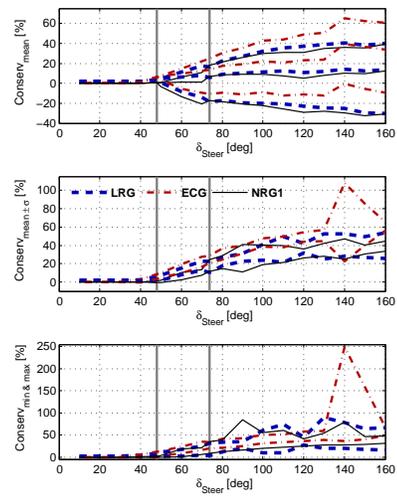}
 \caption{\acp{RG}' effectiveness performance with estimation errors of $ \sigma = 20\% $ about the true roll angle.}
 \label{fig:Results_MonteCarlo_RollError_CtrlPerf_Effectiv}
\end{figure}
\begin{figure}[t]
 \centering
 \includegraphics[width = 2.2in]{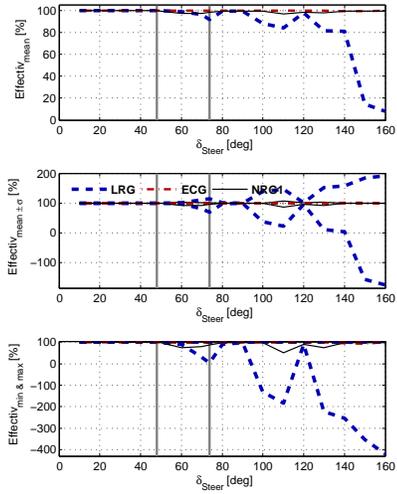}
 \caption{\acp{RG}' conservatism performance with estimation errors of $ \sigma = 20\% $ about the true roll angle.}
 \label{fig:Results_MonteCarlo_RollError_CtrlPerf_ConservRel}
\end{figure}

%========= Figure ===============%
Figures \ref{fig:Results_MonteCarlo_RollError010_CtrlPerf}, \ref{fig:Results_MonteCarlo_RollError_CtrlPerf}, \ref{fig:Results_MonteCarlo_RollError_CtrlPerf_Effectiv}, and \ref{fig:Results_MonteCarlo_RollError_CtrlPerf_ConservRel} illustrate how the errors in the roll angle affect the controllers performance.
%Figures \ref{fig:Results_MonteCarlo_RollError_CtrlPerf}, \ref{fig:Results_MonteCarlo_RollError_CtrlPerf_Effectiv}, \ref{fig:Results_MonteCarlo_RollError_CtrlPerf_ConservRel}, \ref{fig:Results_MonteCarlo_RollError_CtrlPerf_TurnResp}, and \ref{fig:Results_MonteCarlo_RollError_CtrlPerf_CompTime} illustrate how the errors in the roll angle affect the controllers performance.
Figure \ref{fig:Results_MonteCarlo_RollError010_CtrlPerf} shows that even with roll estimation errors up to $ 10 \% $, the effectiveness of all the \acp{RG} is almost unaffected.
Figures \ref{fig:Results_MonteCarlo_RollError_CtrlPerf} and \ref{fig:Results_MonteCarlo_RollError_CtrlPerf_Effectiv} show that the effectiveness of the \ac{ECG} and the \ac{NRG} is almost unaffected even for extremely high roll estimation errors.
The exception is the \ac{LRG} controller.
Its average effectiveness approaches the limit of 0\% for larger reference steering angles, meaning that the wheel could lift an average of $ 45 mm $ ($ 1.8 " $).
Figure \ref{fig:Results_MonteCarlo_RollError_CtrlPerf_Effectiv} indicates that this only happens in the presence of very high roll estimation errors and extremely high steering amplitudes ($ > 140 \; deg $).
%The middle plot shows that for most of the test runs the controller was able to keep the wheels from lifting more than $ 4" $.
%In the extreme the wheels lift was limited to $ 10" $, as is shown by the minimum effectiveness in the bottom plot.
%The most extreme wheel lift situations occur for steering angles of $ 130 degrees $ and above, occasionally allowing wheel lifts above $ 6" $.
%These conditions are extreme in all senses, the roll angle errors are abnormally high, and the steering angles are also quite extreme for a maneuver at $ 50 mph $ ($ 80 kph $).
%========= Figure ===============%
%========= Figure ===============%

%========= Figure ===============%
%========= Figure ===============%
Figure \ref{fig:Results_MonteCarlo_RollError_CtrlPerf_ConservRel} shows that the conservatism is also affected by the estimation errors.
Nevertheless, even with $ 20 \% $ of estimation error in roll, the \acp{RG} conservatism is quite acceptable, and in most test runs it is below $ 50\% $ for the \ac{LRG} and \ac{NRG} and below $ 60\% $ for the \ac{ECG}.
Most importantly, at lower steering angles, even in the most extreme test runs, the controllers conservatism is kept below 12\% for steering angles that would not cause any wheel lift ($ \delta_{SW} < 48 \; deg $) and is kept bellow 35\% for steering angles that would take the wheels to reach the limit wheel lift.
Among the controllers compared, the \ac{ECG} controller seems to be the most sensitive to the roll estimation errors in terms of conservatism.

%%========= Figure ===============%
%\begin{figure}[t]
% \centering
% \includegraphics[width = 2.2in]{RGComp-EstimRoll020-CtrlPerf_TurnResp}
% \caption{\acp{RG}' turning response performance with estimation errors of $ \sigma = 20\% $ about the true roll angle.}
% \label{fig:Results_MonteCarlo_RollError_CtrlPerf_TurnResp}
%\end{figure}
%%========= Figure ===============%
The bottom plots of Figures \ref{fig:Results_MonteCarlo_RollError010_CtrlPerf} and \ref{fig:Results_MonteCarlo_RollError_CtrlPerf} show that the controllers turning response is largely unaffected by the estimation errors.
Unlike in the conservatism metric, the \ac{ECG} seems to be the least affected controller.
%These results also show that the estimation errors do not change much how the car maneuverability is affected by the controller.

%%========= Figure ===============%
%\begin{figure}[t]
% \centering
% \includegraphics[width = 2.2in]{RGComp-EstimRoll020-CtrlPerf_CompTime}
% \caption{\acp{RG}' computation time with estimation errors of $ \sigma = 20\% $ about the true roll angle.}
% \label{fig:Results_MonteCarlo_RollError_CtrlPerf_CompTime}
%\end{figure}
%%========= Figure ===============%
%Figure \ref{fig:Results_MonteCarlo_RollError_CtrlPerf_CompTime} shows that the controllers computation time is only slightly affected by the estimation errors.
%The \ac{NRG} maximum computation time might increase by 10\%, but it is the \ac{ECG} that shows some extreme cases where the computation time can exceed that of the \ac{NRG}.
%The mean and the standard deviation for the \ac{ECG} show that these extreme cases are really exceptions.
%While the \ac{LRG} mean computation time is almost constant at $ 6.5 \times 10^{-3} s $ for all steering angles, the \ac{ECG} peaks at $ 4 \times 10^{-3} s $ for the larger steering angle maneuvers.
%The \ac{LRG} computation time standard deviation is also almost constant at $ 10^{-3} s $ for all steering angles, the \ac{ECG} peaks for the larger steering angle maneuvers and is below $ 5 \times 10^{-3} s $ for almost all steering angles.

%========= Figure ===============%
\begin{figure}[t]
 \centering
 \includegraphics[width = 2.2in]{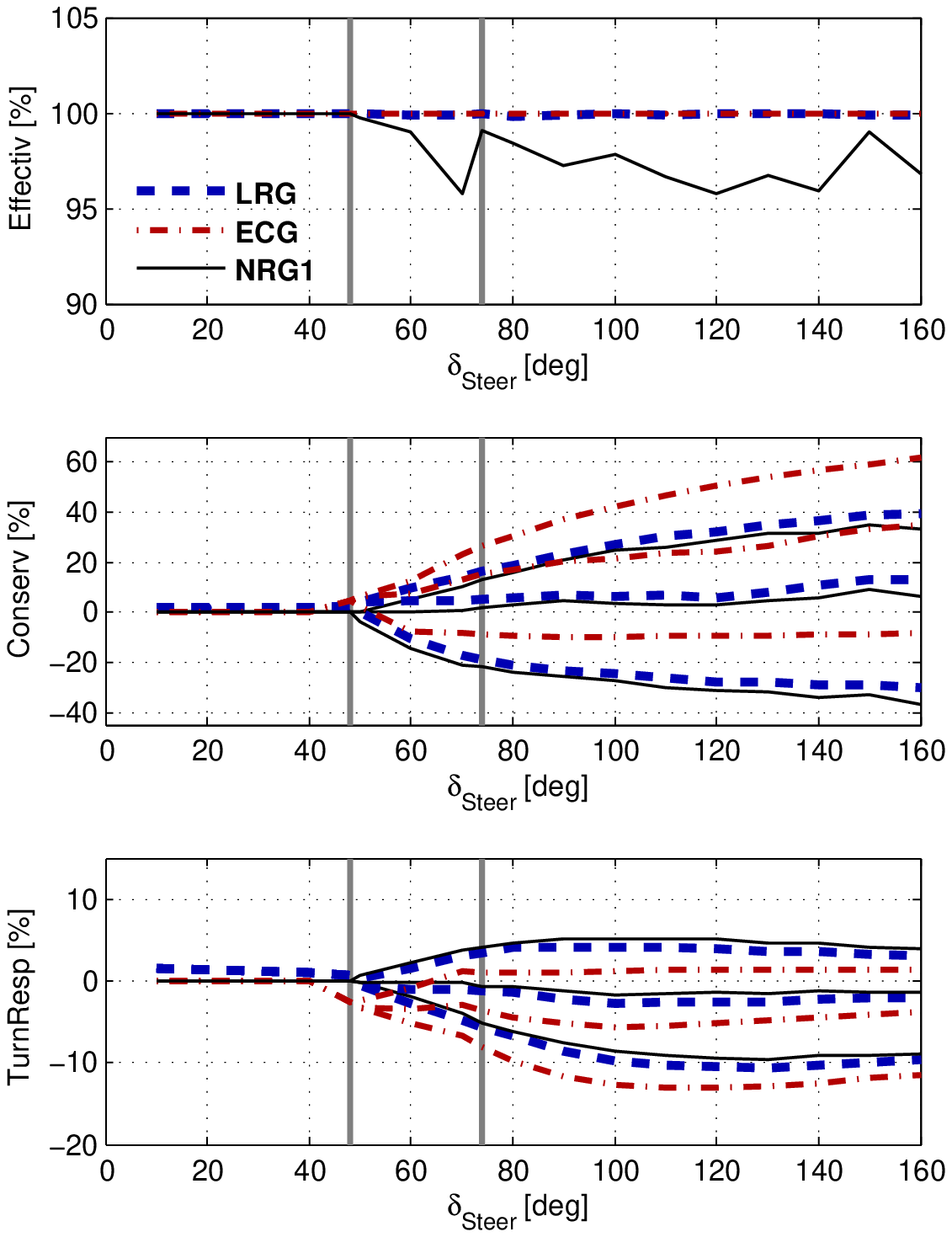}
 \caption{\acp{RG}' performance with estimation errors of $ \sigma = 20\% $ about the true roll rate.}
 \label{fig:Results_MonteCarlo_RollRtError_CtrlPerf}
\end{figure}
%========= Figure ===============%
%========= Figure ===============%
\begin{figure}[t]
 \centering
 \includegraphics[width = 2.2in]{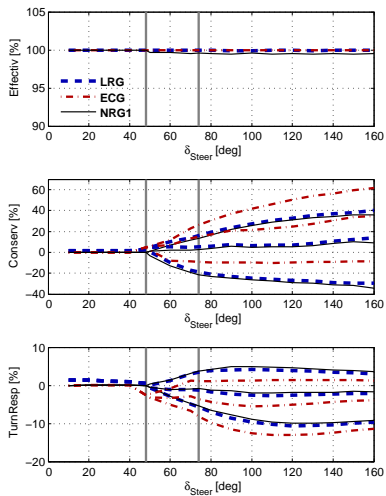}
 \caption{\acp{RG}' performance with estimation errors of $ \sigma = 20\% $ about the true side-slip angle.}
 \label{fig:Results_MonteCarlo_SwayError_CtrlPerf}
\end{figure}
%========= Figure ===============%
%========= Figure ===============%
\begin{figure}[t]
 \centering
 \includegraphics[width = 2.2in]{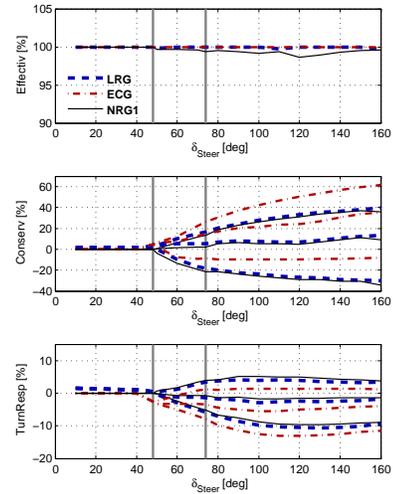}
 \caption{\acp{RG}' performance with estimation errors of $ \sigma = 20\% $ about the true turn rate.}
 \label{fig:Results_MonteCarlo_YawRtError_CtrlPerf}
\end{figure}
%========= Figure ===============%
Figures \ref{fig:Results_MonteCarlo_RollRtError_CtrlPerf}, \ref{fig:Results_MonteCarlo_SwayError_CtrlPerf}, and \ref{fig:Results_MonteCarlo_YawRtError_CtrlPerf} show that the controllers' performance is only slightly affected by roll rate estimation errors and is not visibly affected by the side-slip and turn rate errors, even for the high estimation errors ($ \sigma = 20\% $).

%========= Figure ===============%
\begin{figure}[t]
 \centering
 \includegraphics[width = 2.2in]{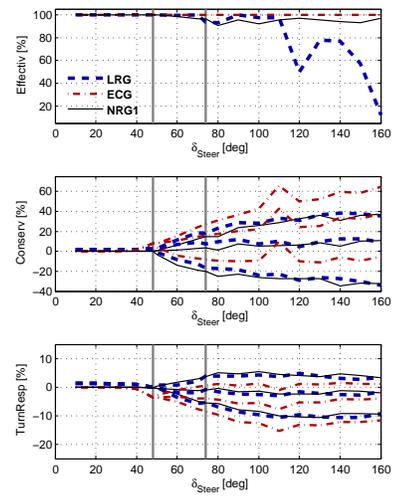}
 \caption{\acp{RG}' performance with estimation errors of $ \sigma = 20\% $ about the true roll angle and true roll rate.}
 \label{fig:Results_MonteCarlo_RollRollRtError_CtrlPerf}
\end{figure}
%========= Figure ===============%
Figure \ref{fig:Results_MonteCarlo_RollRollRtError_CtrlPerf} shows that the effect on the controllers performance from a combination of roll and roll rate estimation errors is very similar to that of just the roll estimation errors.
These results show that the controllers are most sensitive to the errors in the roll angle.
Nevertheless, the controllers average response is adequate, even with estimation errors with a standard deviation of 20\%, which is an extremely poor estimation performance.

%% file: Conclusions.tex
\section{Conclusions and Future Work}
\label{sec:Conclusions}

%=====================================================================================%
\subsection{Conclusions}

We have presented several \acf{RG} designs for vehicle rollover avoidance using active steering commands.
We implemented three types of \aclp{RG}: a \acf{LRG}, an \acf{ECG}, and a \acf{NRG}.
The goal of the \aclp{RG} is to enforce \acf{LTR} constraints.
The \aclp{RG} predict the vehicle trajectory in response to a reference steering command, from the driver, to check if it respects the \acf{LTR} constraints.
The \ac{LRG} and the \ac{ECG} use a linear model to check the safety of the steering command.
The \ac{NRG} uses a nonlinear model to achieve the same goal.
%In case the driver steering command is deemed unsafe, the \ac{ECG} uses an LQ optimizer
The controllers were tested with a nonlinear simulation model to check their performance with realistic vehicle dynamics, that are highly nonlinear for large steering angles.
The nonlinearity causes the standard versions of the \ac{LRG} and the \ac{ECG} to be too conservative.
We have presented several methods to compensate for such nonlinearities.

To evaluate the controllers we have defined three performance metrics: effectiveness, conservatism, and turning response.
The effectiveness characterizes how well the constraints are enforced by the \ac{RG}.
The conservatism and the turning response characterize if the controller is too intrusive or not, by measuring how well the \ac{RG} command and respective vehicle trajectory adhere to the driver command and desired trajectory.
An additional evaluation metric is the command computation time, that characterizes the controller computational load.
The \ac{NRG} provides the best performance in terms of effectiveness, conservatism, and turning response, but it is slower to compute.
The simulation results show that \ac{LRG} with the nonlinear compensations (\emph{nonlinear difference} and \emph{\ac{MPL}}) provides the best balance between all the metrics.
It has a low computational load, while showing very high constraint enforcement effectiveness and generating commands with a conservatism almost as low as the \ac{NRG}.
The simulation results also show that the \aclp{RG}' performance is most sensitive to the roll estimation errors, but that even with very high estimation errors ($ \sigma = 20\% $) the \aclp{RG} can still enforce the constraints effectively.

%=====================================================================================%

\section{Further Extensions}

Currently we are working to extend the current approach to incorporate differential braking and active suspension commands.
It is important to understand the performance limits in rollover avoidance for each individual command.
It is also important to understand how the different commands can be combined to provide the most effective and least intrusive rollover avoidance intervention.

%To improve the \ac{ECG} performance, a constraint relaxation method can be developed (sec.~\ref{sec:Ctrl_Feasibility_ConstrRelax}).
Research is also needed to determine the slowest acceptable control update rate, \emph{i.e.}, to characterize how slower update rates degrade the control performance and the driver handling perception.

This research shows that the presented \aclp{RG} can cope with a great deal of estimation errors.
Further research should address the state estimation methodology, given the limited sensing capabilities in standard cars.
With such methodology, a better estimation error model should be integrated with the vehicle simulation to verify the \aclp{RG} performance with a realistic estimation error model.
The effects of vehicle and road uncertainties on the estimation and control performance also need to be studied, in order to understand how the system will perform in the range of operating conditions in which the real vehicles will operate.